\documentclass[10pt,preprint]{aastex}

\newcommand{\myemail}{mahmoud.parvizi@vanderbilt.edu}

\shorttitle{Eclipsing Binary Factory}
\shortauthors{Parvizi et al.}

\usepackage{color}
\usepackage[normalem]{ulem}

\begin{document}

%\title{Searching the \textit{Kepler} Field with the Eclipsing Binary Factory Pipeline}
\title{The EB Factory Project. II. Validation with the \textit{Kepler} Field in 
Preparation for K2 and TESS}

\author{Mahmoud Parvizi\altaffilmark{1,2}, 
Martin Paegert\altaffilmark{1},
Keivan G.\ Stassun\altaffilmark{1,3}}
\altaffiltext{1}{Department of Physics and Astronomy, Vanderbilt University,
    VU Station B 1807, Nashville, TN 37235, USA \\ \myemail}
\altaffiltext{2}{Department of Physics and Astronomy, Austin Peay State University,
    601 College St, Clarksville, TN 37044, USA}
\altaffiltext{3}{Department of Physics, Fisk University, 1000 17th Ave. N., Nashville, TN 37208, USA}

\begin{abstract}
Large repositories of high precision light curve data, such as the \textit{Kepler} data set, provide the opportunity to identify astrophysically important eclipsing binary (EB) systems in large quantities. However, the rate of classical ``by eye" human analysis restricts complete and efficient mining of EBs from these data using classical techniques. To prepare for mining EBs from the upcoming K2 mission as well as other current missions, we developed an automated end-to-end computational pipeline --- the Eclipsing Binary Factory (EBF) --- that automatically identifies EBs and classifies them into morphological types. The EBF has been previously tested on ground-based light curves. To assess the performance of the EBF in the context of space-based data, we apply the EBF to the full set of light curves in the \textit{Kepler} ``Q3'' Data Release. We compare the EBs identified from this automated approach against the human generated \textit{Kepler} EB Catalog of $\sim2,600$ EBs. When we require EB classification with $\ge$ 90\% confidence, we find that the EBF correctly identifies and classifies eclipsing contact (EC), eclipsing semi-detached (ESD), and eclipsing detached (ED) systems with a false positive rate of only 4\%, 4\%, and 8\%, while complete to 64\%, 46\%, and 32\% respectively. When classification confidence is relaxed, the EBF identifies and classifies ECs, ESDs, and EDs with a slightly higher false positive rate of 6\%, 16\%, and 8\%, 
while much more complete to 86\%, 74\%, and 62\% respectively. Through our processing of the entire \textit{Kepler} ``Q3'' dataset, we also identify 68 new candidate EBs that may have been missed by the human generated \textit{Kepler} EB Catalog. We discuss the EBF's potential application to light curve classification for periodic variable stars more generally for current and upcoming surveys like K2 and the Transiting Exoplanet Survey Satellite.
\end{abstract}


\section{Introduction}

Periodic variable stars, in particular eclipsing binary (EB) star systems, continue to be of central importance to a variety of applications in stellar astrophysics. EBs are crucial astrophysical benchmarks for stellar evolution theory as they can provide fundamental physical parameters of stars with very high precision and accuracy. Fundamental issues, from the stellar mass-radius relation using EBs in nearby clusters and the field \citep[e.g.,][]{Lacy1977}, to the extragalactic distance scale using EBs in the Large Magellanic \citep{Pietrzynski2013} and Small Magellanic Cloud \citep{Graczyk2014}, rely on the discovery and analysis of statistically significant samples of EBs and/or of rare but astrophysically important EBs. Indeed, EBs in which at least one component is of very low mass, in a scarce evolutionary phase, or itself a periodic variable are astrophysically very interesting and useful, yet rare and thus only small numbers of these benchmark systems are found in current catalogs. For example, there is only one known EB system comprising two brown dwarfs \citep{Stassun2006,Stassun2007}, and only a couple EBs containing a Classical Cepheid pulsators are currently known \citep[e.g.,][]{Pietrzynski2010}. Repositories of large time-series photometric survey data, such as the \textit{Kepler} archive, provide an ideal sample from which to identify periodic variables including these interesting EBs for focused followup investigation. However, constraints on the rate of human analysis make mining this data using classical techniques prohibitive. 

An intelligent data pipeline (IDP) that contains an automated light curve classifier to replace the human bottleneck is then necessary for the future pace of large survey data rates. Ideally such an IDP would be capable of automatically finding EBs in light curve data sets and then accurately classifying these EBs by morphological type (e.g., eclipsing  contact, eclipsing semi-detached, eclipsing detached) to enable focused followup analyses. Several algorithms have been developed for this purpose (e.g., \citet{Wyrzykowski2003,Devor2008,Graczyk2011,Devinney2012} these are discussed further in Sec.~\ref{DIS}), yet a fully-automated pipeline remains an open task.

An automated classification pipeline with parameters adaptable to multiple time series photometric surveys would be immediately applicable to the \textit{Kepler} data set \citep{Christiansen2013} as well as the follow-on K2 mission \citep{Howell2014} and would be well suited to the upcoming Transiting Exoplanet Survey Satellite (TESS) \citep{Ricker2014}. In addition, the development of the Large Synoptic Survey Telescope (LSST) contains a requirement that data processing enable a fast and efficient response to transient sources (i.e., automated identification of variable stars and astrophysically interesting binaries) with a robust and accurate preliminary classification \citep{Ivezic2011}. In short, there is a large need currently and into the near future for methods that can reliably process large quantities of light curve data and automatically identify astrophysical objects of interest such as EBs with minimal human intervention.

So motivated, especially the immediate applicability of an automated pipeline for the study of EBs in K2 \citep{Welsh2013}, in this paper we present a fully automated, adaptive, end-to-end computational pipeline --- the Eclipsing Binary Factory (EBF). The EBF first identifies EBs, and then morphologically classifies them as either eclipsing contact (EC), eclipsing semi-detached (ESD), or eclipsing detached (ED). The classification module of the EBF pipeline was described in the first paper in this series \citep{Paegert2014}; here we add the remaining computational modules for a complete automated pipeline. In this paper we also then measure the accuracy and completeness of our EBF generated catalog of \textit{Kepler} EBs against the manually constructed \textit{Kepler} EB Catalog \citep{Prsa2011,Slawson2011}. 

The EBF takes a modular approach to automatically process large volumes of survey data into patterns suitable for recognition by an integrated, fully-trained artificial neural network (ANN). This modular pipeline is designed to accept as input the direct archival data from a time-series photometric survey where each module's parameters are adaptable to the specific characteristics of the data (e.g., photometric precision, data collection cadence, flux measurement uncertainty) and define the output options to produce the ANN generated catalog of EB classifications (i.e., EC, ESD, or ED). The computational efficiency of the ANN classifier, $O(2N)$ where $N$ is the number of neurons in the neural network, is quite fast---of order $0.8$ seconds for classification of $\sim50,000$ light curves using a standard workstation. This then alleviates the current bottleneck by limiting followup human analysis to just the light curves automatically classified as EBs. In addition, the portability of this modular software pipeline makes the EBF a solution for the automated probabilistic identification and morphological classification of EBs in other large repositories of time-series, photometric data currently underway or planned for the near future.

The \textit{Kepler} mission monitored $\sim200,000$ targets brighter than the magnitude limit $Kp \sim16$ in the \textit{Kepler} spacecraft's field of view (FoV). \textit{Kepler} captured and summed these images as long cadence, 29.4 $minutes$, data files \citep{Jenkins2010} that contain one quarter, or $\sim$ 90 $days$, of observations across the FoV for sixteen quarters until reaction wheel failure placed the instrument into safe mode. The complete scientific data set, archived in MAST at the Space Telescope Science Institute, provides simple aperture photometry (SAP) with the photometric precision necessary to detect planetary transits \citep{Koch2010} and astero-seismological signals \citep{Gilliland2010} from solar-like stars; about 20 $ppm$ for $12^{th}$ magnitude G2V stars for a 6.5 $hour$ integration \citep{VanCleve2009}. As such, this data set provides both the appropriate sample size and precision for use in testing the EBF pipeline. In addition, the \textit{Kepler} Eclipsing Binary Catalog Version 3 (KEBC3) of $\sim2,600$ manually classified EBs and a corresponding catalog of manually identified false positives in the \textit{Kepler} FoV \citep{Prsa2011,Slawson2011} provides a benchmark of the EBF's classification accuracy (i.e., false positive rate) and completeness (i.e., false negative rate). Moreover, full processing of the Kepler dataset with the EBF automated pipeline provides the opportunity to uncover some additional EBs that may have been missed by the human generated KEBC3.

We demonstrate the efficacy of the EBF by automatically identifying and morphologically classifying EBs in the $\sim165,000$ \textit{Kepler} Quarter 3 (``Q3'') SAP light curves. In Section \ref{MET} of this paper we describe the procedures and methods of each EBF module. Section \ref{RES} presents the results of the EBF applied to the Kepler ``Q3'' light curves. Section \ref{resp} benchmarks the pre-classification data processing modules against the metrics of the manually corrected and phased light curves in the KEBC3 while Section \ref{resc} focuses on the EBF pipeline's false positives and false negatives as a measure of classification accuracy and completeness. Finally, in Section \ref{DIS} we discuss the future expansion of the EBF to include the automated identification and classification of other periodic variable stars such as $\delta$ Cepheid fundamental mode (DCEP-FU) and first overtones (DCEP-FO), $\delta$ Scuti (DSCT), RR Lyrae $ab$ (RRAB) and $c$ (RRC), and Mira (MIRA) as well the EBF's applicability to other large time-series, photometric surveys. A set of newly identified candidate EBs that may have been missed by the human generated \textit{Kepler} EB Catalog is provided in Appendix \ref{APP}. We conclude with a brief summary in Section \ref{CON}.

\section{Methods: The Eclipsing Binary Factory Pipeline \label{MET}}

\begin{figure}[!hb]
\begin{center}
\includegraphics[scale=0.4]{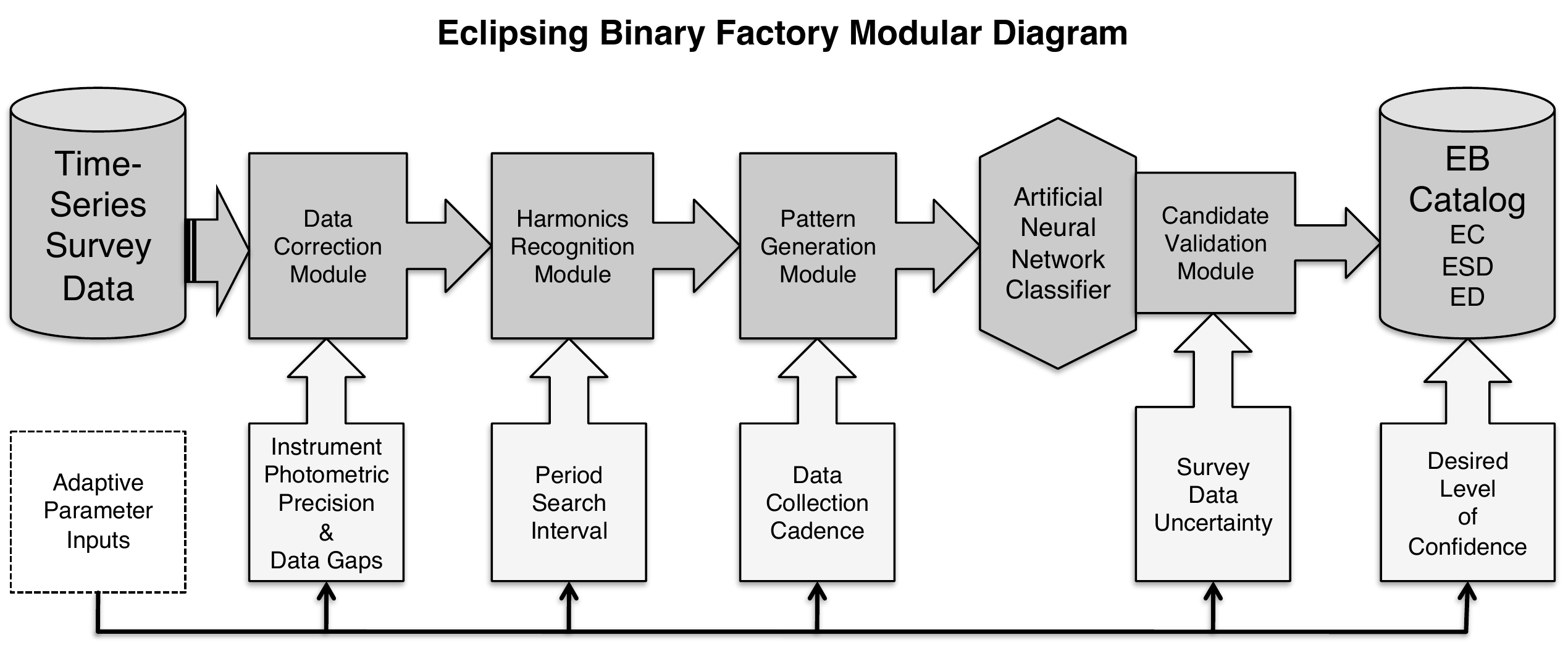}
\caption{\footnotesize{The fully automated and adaptive Eclipsing Binary Factory pipeline, diagrammed by module, for processing time-series photometric data from large astronomical surveys in order to automatically identify EBs and sub-classify the system's morphology as eclipsing contact (EC), eclipsing semi-detached (ESD), or eclipsing detached (ED) by an artificial neural network classifier.}}
\label{Figure1}
\end{center}
\end{figure}

The EBF Pipeline, diagrammed in Figure \ref{Figure1}, is a compilation of five fully automated, adaptable software modules that process light curves from large astronomical surveys in order to automatically identify EBs and classify the each EB's morphology. In this section we describe the parameters with which systematics are removed, periodic variable stars are identified, phased light curves are represented as ANN recognizable patterns, and constrain how EB candidate classifications are tuned and validated as a function of the survey instrument's photometric precision, the data set duration and cadence, the flux measurement uncertainty, and the desired level of classification confidence, respectively. The procedure of each EBF module is described below in turn, then we test the overall pipeline performance by the rate of correctly classified systems and the completeness of the EBF classified sample as compared to the KEBC3 benchmark sample classified by human analysis. We use the \textit{Kepler} ``Q3" long cadence data files as described in the \textit{Kepler} Data Release 4 Notes KSCI-199044-001 \citep{VanCleve2010}, to tool and test the EBF pipeline.

\subsection{Data Correction Module \label{dcm}}

Corrections to the raw photometry collected from large survey instruments are generally necessary to remove systematic artifacts from the data. Though most surveys include their own data correction processes, there is no guarantee that the method employed preserves the necessary signal from which the EBF will identify and classify EBs. In the case of the \textit{Kepler} ``Q3'' light curves corrections to the SAP flux are necessary to remove systematics such as differential velocity aberration, thermal gradients across the spacecraft, and pointing variations \citep{Kinemuchi2012}. Of course, the \textit{Kepler} archive data reduction pipeline removes systematics in the ``Q3'' SAP data to produce pre-search conditioned data (PCD) using the method of cotrending basis vectors; however, PCD may lack the astrophysical signatures of variables whose harmonic thresholds were not met \citep{McQuillan2012}. Hence, we employ a data correction module (DCM) to remove trending artifacts while preserving the necessary signal as well as to normalize the flux in order to present the remainder of the pipeline with a standardized, detrended light curve. The DCM step is applied to all of the input light curves (i.e., no light curves are filtered out at this stage).

\begin{figure}[!ht]
\begin{center}
\plottwo{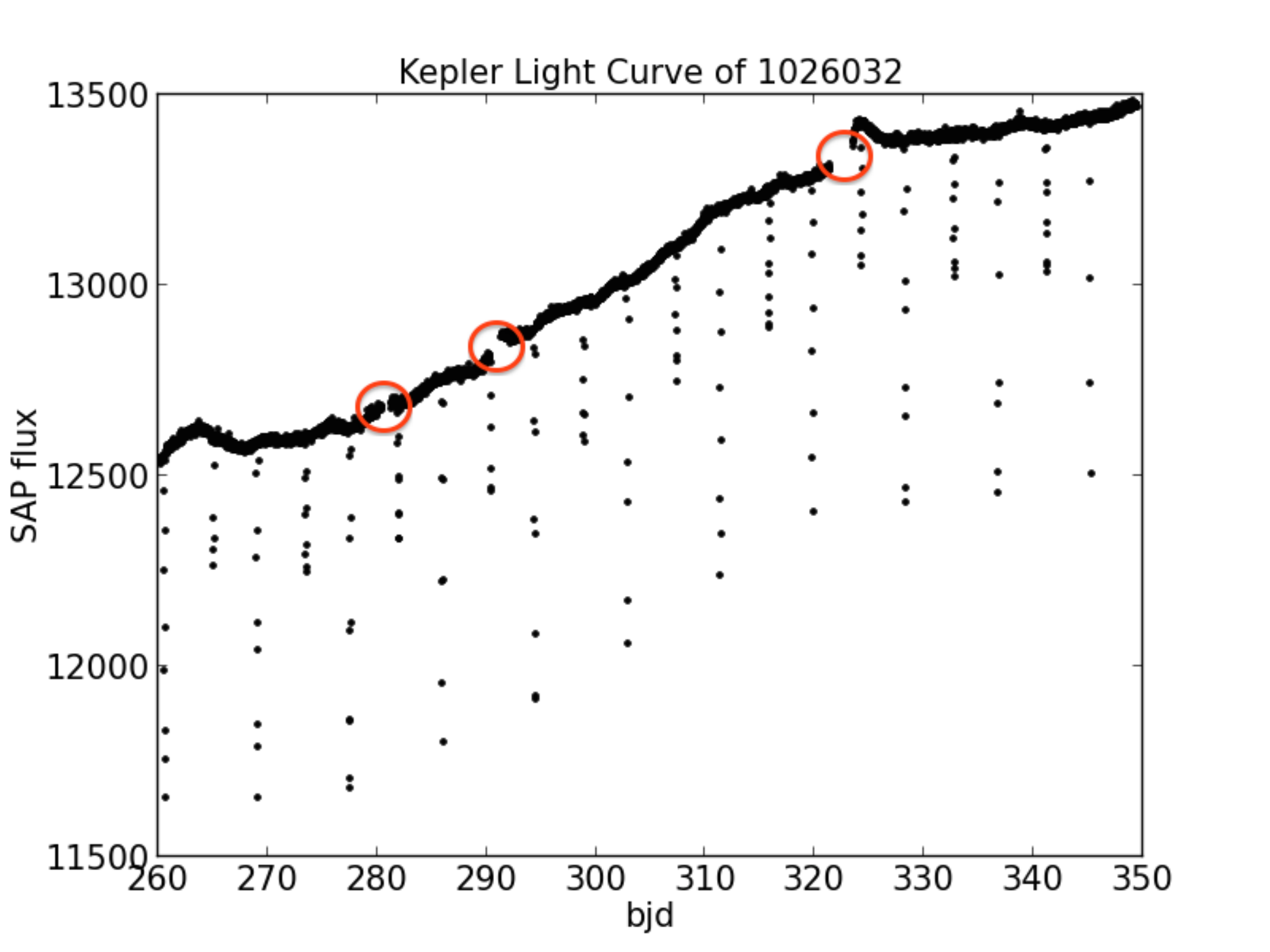}{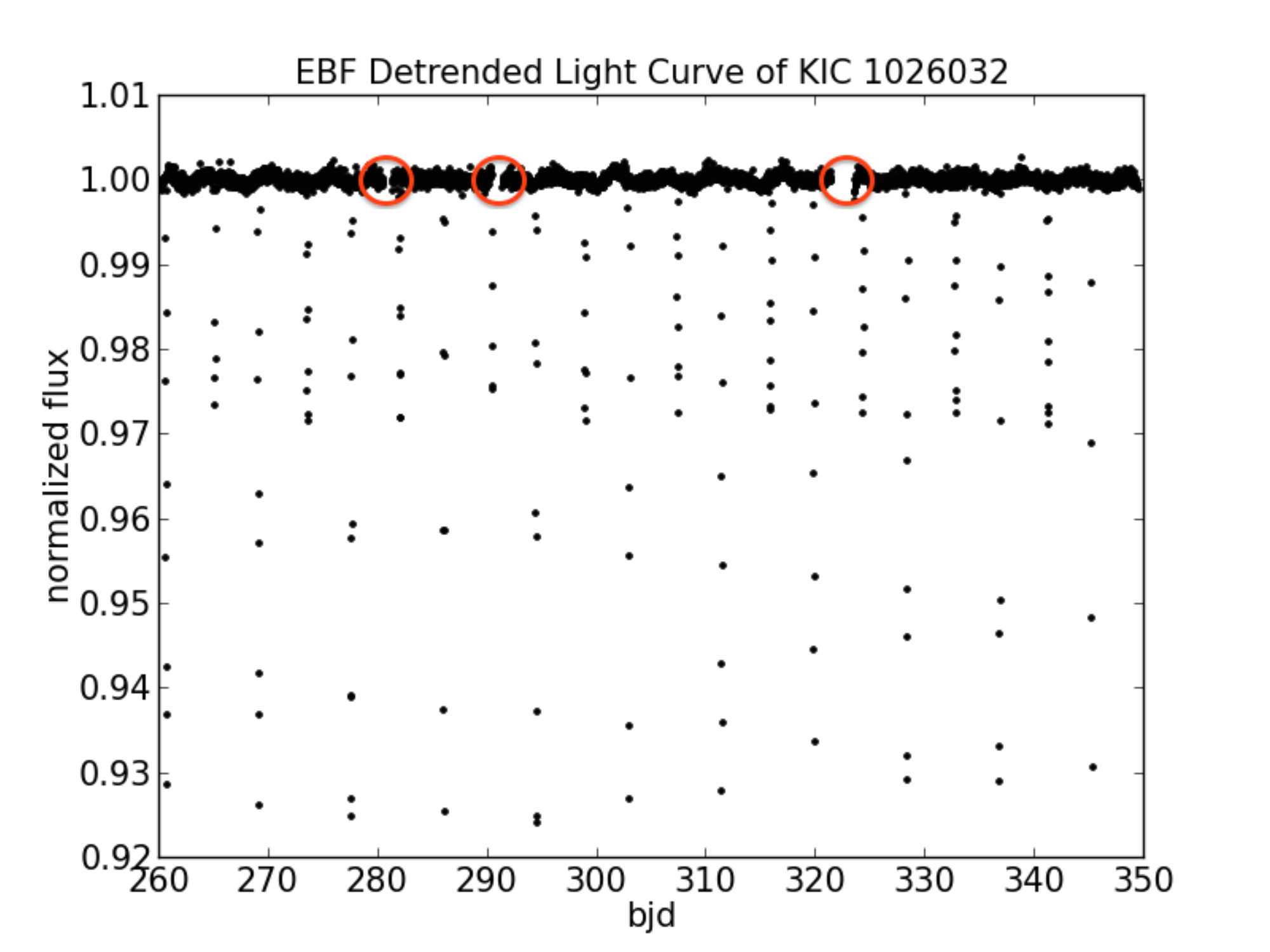}
\caption{\footnotesize{\textbf{Left:} The \textit{Kepler} ``Q3'' SAP light curve for target KIC 1026032. \textbf{Right:} The EBF detrended, normalized light curve for target KIC 1026032 where systematics were removed by the data correction module using the sigma-clipping algorithm of \citet{Slawson2011} for asymmetric sigma-clipping thresholds $(-1\sigma, 3\sigma)$ with a least squares fit to the $10^{th}$ order \textit{Legendre} polynomial. The red circles indicate data gaps that are accounted for by the pipelin.}}
\label{Figure2}
\end{center}
\end{figure}

The DCM accepts as input the time-series flux with the associated uncertainty of each data point and detrends the flux via the sigma-clipping algorithm of \citet{Slawson2011}; where data points outside a standard deviation interval of a least squares \textit{Legendre} polynomial fit to order $l$, expressed by the generating function
\begin{equation}
P_l(x) = \frac{1}{2^ll!} \frac{d^l}{dx^l}[(x^2-1)^l],
\end{equation}
are iteratively discarded until there are no remaining data points outside the interval. For each data point input, the DCM returns a point containing a normalized flux value with a new uncertainty. The EBF's implementation of this method allows the parameters of the module to be set to the specific characteristics of a data set by defining the sigma clipping thresholds and \textit{Legendre} polynomial order. Additionally, the module allows for the input of the time intervals of known breaks in the light curves due to instrument operations. In this paper we detrend the \textit{Kepler} ``Q3'' SAP light curves using asymmetric sigma-clipping thresholds of $(-1\sigma, 3\sigma)$ with the $10^{th}$ order \textit{Legendre} polynomial
\begin{equation}
P_{10}(x) = 256^{-1} (46,189x^{10} - 109,395x^{8} + 90,090x^{6} - 30,030x^{4} - 3,465x^{2} - 63)       
\end{equation}
and known breaks in the ``Q3'' light curves beginning at BJD 281.0, 291.0, and 322.5. Figure \ref{Figure2} shows the uncorrected \textit{Kepler} ``Q3'' SAP light curve for a target and the EBF detrended, normalized light curve where systematics were removed by the DCM and the red circles indicate data gaps given by the aforementioned breaks.

%\pagebreak

\subsection{Harmonics Recognition Module \label{hrm}}

The EBF pipeline is currently optimized for identification and classification of EBs, which exhibit periodically varying light curves. Therefore the precursors to morphological classification of EBs via pattern recognition are the identification of a target's periodic variability and a subsequent phase folded light curve. Hence, the detrended normalized light curve is next processed by the harmonics recognition module (HRM) where periodic variability is probed by the analysis of variance (AoV) method \citep{Schwarz1989,Devor2005} included in the VARTOOLS1.202 package \citep{Hartman2008}. The HRM takes as input all of the normalized light curves output by the preceding DCM, but outputs only those light curves that are deemed to exhibit truly periodic behavior (of any kind, i.e., the output light curves from the HRM stage include EBs but also any other type of periodic variable).

Two separate searches for the first three harmonics, each with a coarse sub-sample setting of 0.1 and a fine-tune setting of 0.01, are conducted over the entire data set. The first search separates the data into five bins and the second uses fifty bins such that skewed harmonics from bin size related flux variance is minimized. We then compare the strongest harmonic signal from both periodograms and, in the event of disagreement between the primary periods, assign the period (along with all other associated AoV statistics) from the periodogram with the smallest false alarm probability (FAP). 

To eliminate light curves that are unlikely to exhibit truly significant periodic variability, we use the AoV calculated signal to noise ratio (SNR). Specifically, periodograms with a SNR $\le$ 100 are rejected by the module while the remaining light curves, now identified as candidate periodic variables, are accepted. These accepted candidates' light curves are phased by the standard equation 
\begin{equation}
\rho = \frac{t-t_{\emptyset}}{ p} - int \bigg( \frac{t-t_{\emptyset}}{p} \bigg)
\end{equation}
where $\rho$ is the phase of an associated time value, $t$, for a given reference time, $t_{\emptyset}$, and the assigned period, $p$. 

For an EB, we want the reference or zero phase $\rho_\emptyset$, where $t=t_{\emptyset}$, to be associated with the primary eclipse. However, not all of the periodic variables identified by the HRM will be EBs; there will be other types of periodic variables where the most physically
meaningful reference phase is not an eclipse but rather a flux maximum. Therefore, to maintain generality, we associate $\rho_\emptyset$ with either the minimum or maximum flux, dependent upon whether the light curve spends most of its time above or below the median flux. Most true EBs will have flux points in the ``high" state and will spend
a minority of the time in eclipse. Here the determination is made by the value of the flux ratio (FR) as described by \citet{Coughlin2011}: 
\begin{equation}
FR = \frac{({\rm Maximum Flux} - {\rm Median Flux})}{({\rm Maximum Flux} - {\rm Minimum Flux})}.
\end{equation}
By definition, the FR has a value on the interval (0,1). For reference, a perfect sinusoid will have FR = 0.5. We assign $\rho_\emptyset$ to the flux minimum, indicative of an EB, for a FR less than a defined threshold, and a flux maximum assigned at $\rho_\emptyset$ otherwise.

\begin{figure}[ht]
\plottwo{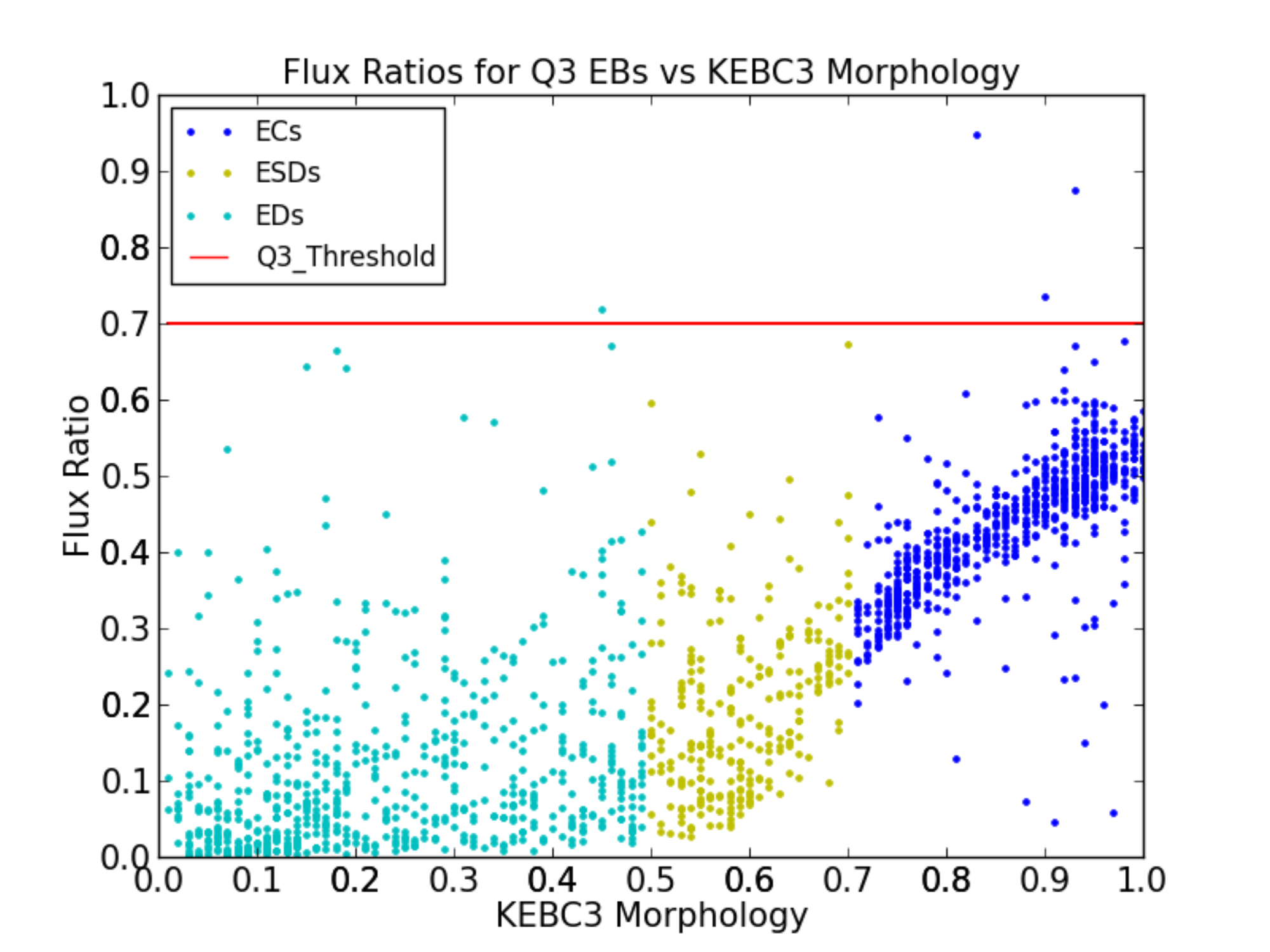}{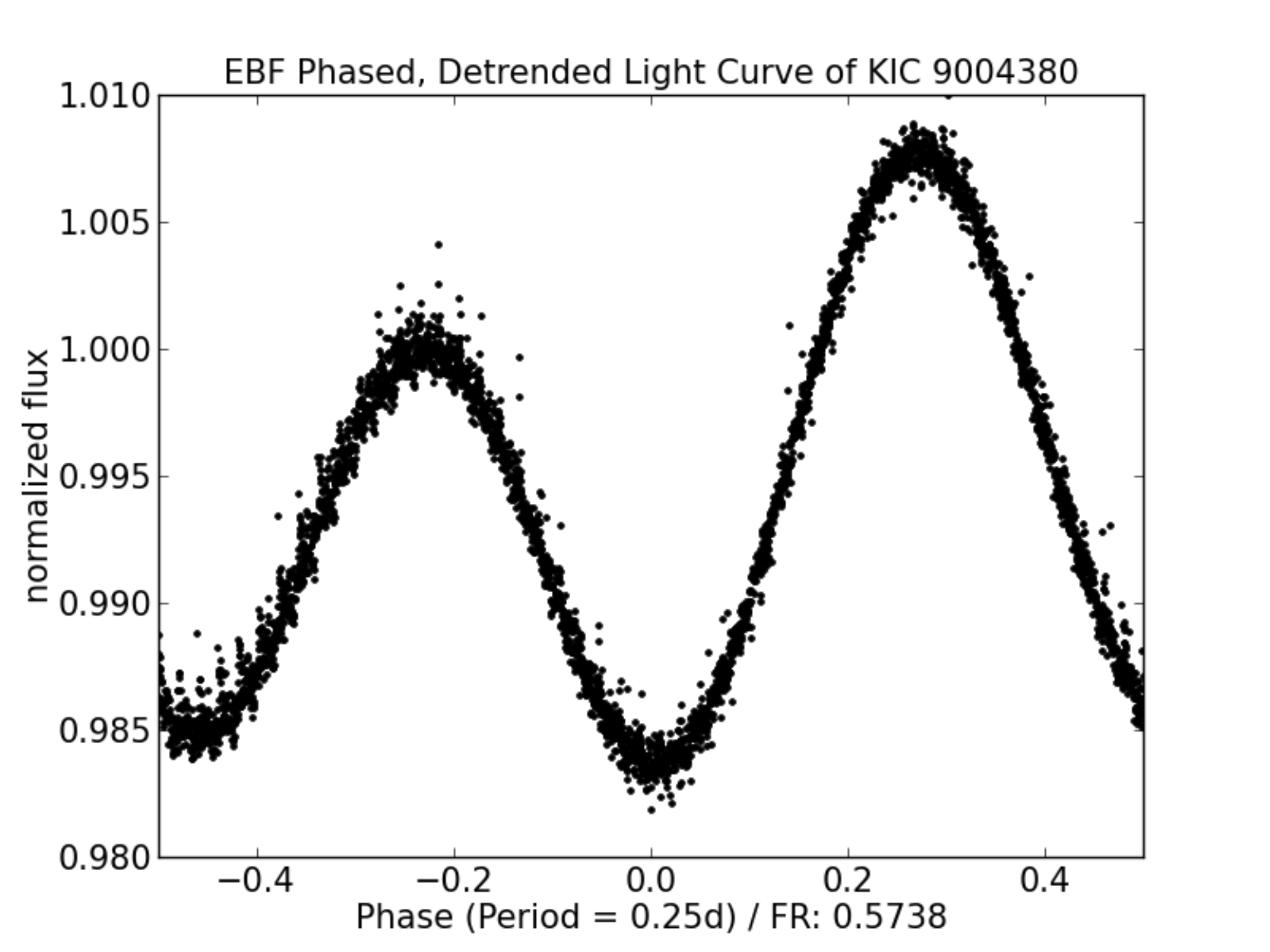}
\caption{\footnotesize{\textbf{Left:} The flux ratio dispersion in the KEBC3 vs the cataloged morphology, where the cyan dots are EDs, the yellow dots are ESDs, and the blue dots are ECs.The red line marks the KEBC3 threshold for imperfect sinusoidal light curves. \textbf{Right:} EBF phased, detrended light curve of KIC 9004380 as an example of an imperfect sinusoidal EC light curve where $0. 5<FR<0.7$.}}
\label{Figure3}
\end{figure}

\begin{figure}[ht]
\plottwo{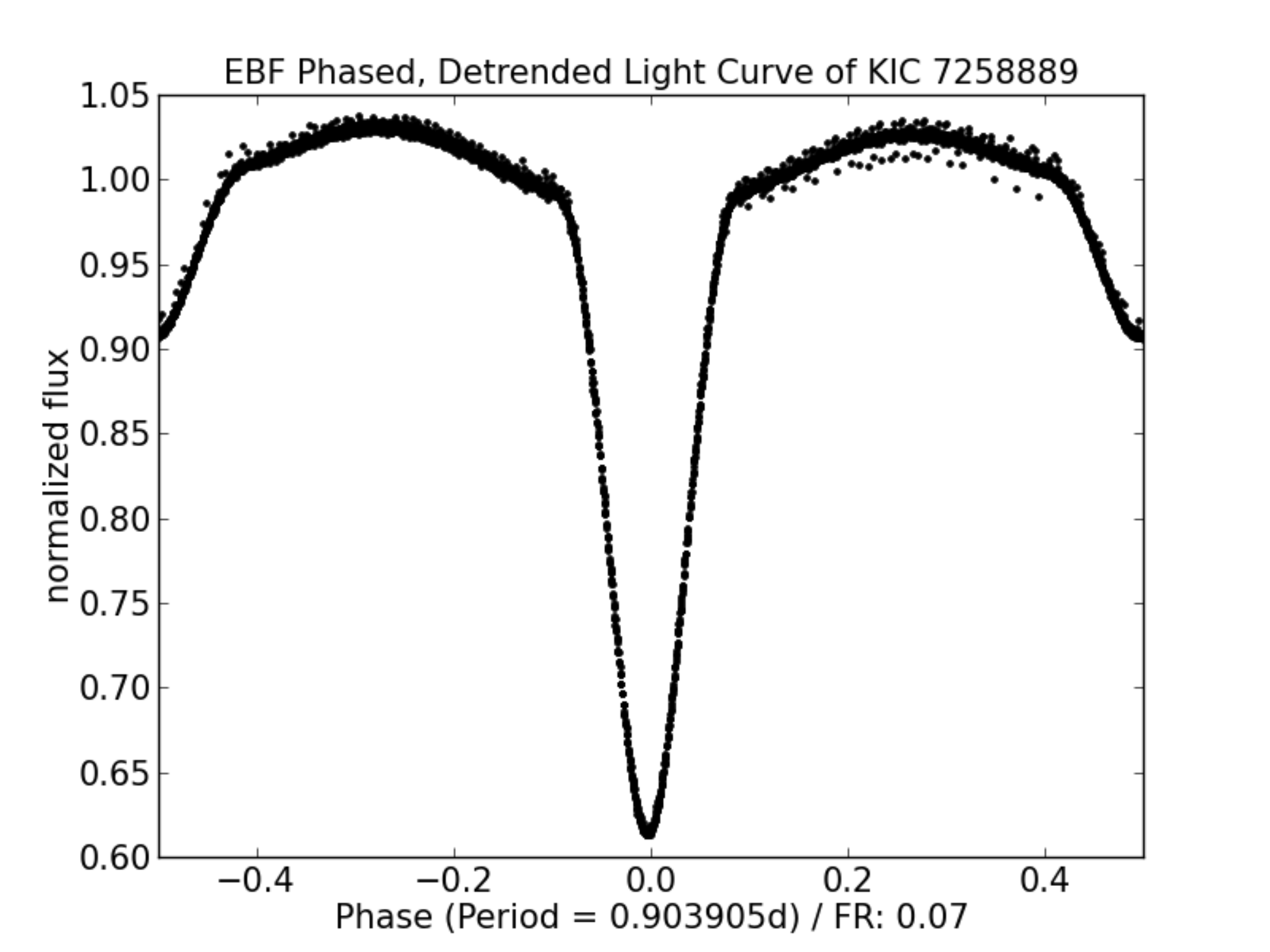}{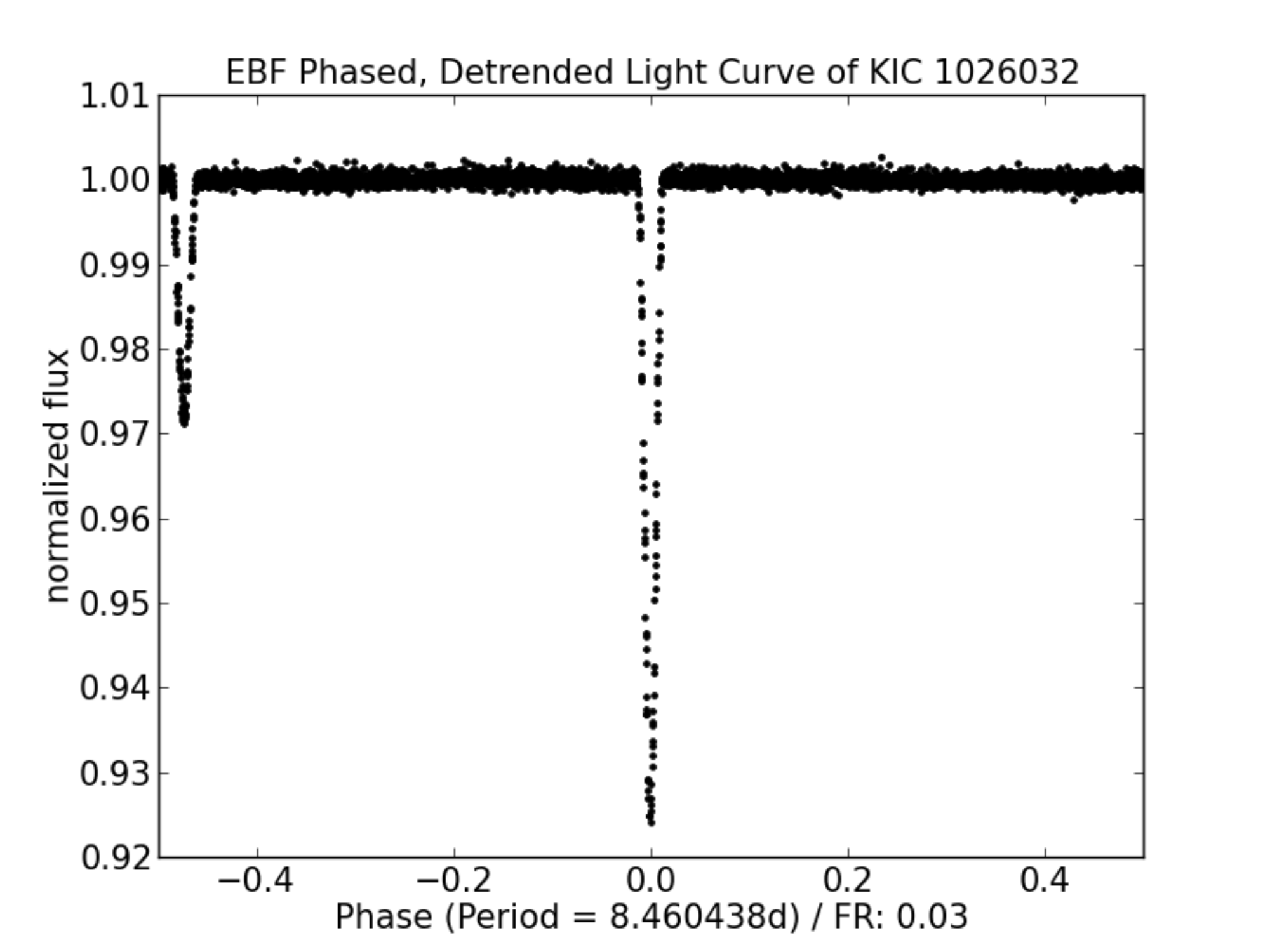}
\caption{\footnotesize{\textbf{Left:} EBF phased, detrended light curve of KIC 7258889 as an example of an ESD light curve with a $FR<0.5$. \textbf{Right:} EBF phased, detrended light curve of KIC 1026032 as an example of an ED light curve with a $FR<0.5$.}}
\label{Figure4}
\end{figure}

The HRM's parameters that are adaptable to the data set are the period search interval, based on the duration of the survey data, and the FR threshold. The latter is based on the tolerance for ECs, which in general are not perfect sinusoids and whose FR may be greater than 0.5 if they spend more than half the time above the median flux. As we limited the \textit{Kepler} input files to ``Q3'' ($\sim$90 $days$), the harmonics search was limited to a period range of 0.11 to 20.1 $days$ in order to ensure at least three primary eclipses would be present in each light curve. In addition, we set the FR threshold to 0.7 based on a simple analysis of KEBC3 FRs in order to include those ECs whose light curves are not perfectly sinusoidal. Figure \ref{Figure3} shows a plot of the FR distribution in the KEBC3 versus cataloged morphology, and a ``Q3'' target as an example of an imperfect sinusoidal phased EC light curve with a $FR>0.5$. An example of a phased ESD light curve, and an example of a phased ED light curve, each with a $FR<0.5$, are shown in Figure \ref{Figure4}. 

\subsection{Pattern Generation Module \label{pgm}}

As described in Section \ref{ann}, the EBF's classification of the phased light curves for the candidate periodic variables output from the HRM is dependent upon the ANN evaluating these light curves as linearly separable patterns. In its simplest form linearly separable means two sets of points in a plane may be divided by at least one line such that each data point belonging to one set is on one side of the line and each data point belonging to the second set is on the other side of that line. When generalized to an n-dimensional space, linear separability is formalized such that every point $x \in \mathbf{X_0}$ satisfies $\sum\limits_{i=1}^n w_ix_i > k$ and every point $x \in \mathbf{X_1}$ satisfies $\sum\limits_{i=1}^n w_ix_i < k$, where $\mathbf{X_0}$ and $\mathbf{X_1}$ are two sets of points in an $n-$dimensional \textit{Euclidean} space and $x_i$ is the $i^{th}$ component of $\mathbf{X_j}$ for $n+1$ real numbers $w_1$, $w_2$ ..., $w_n$, $k$. The pattern generation module (PGM) therefore represents the phased light curves via piecewise smooth polynomial chains $\mathbf{X_j}$ to pass to the ANN as linearly separable patterns. 

Since EBF efficiency goes as $O(N^2)$ where $N$ is the number of elements in $\mathbf{X_j}$, it becomes computationally expensive to process light curves possessing on the order of $10^3$ data points. Therefore, we condense the light curve by phase binning down to the order $10^2$ data points over the phase interval $(-0.5, 0.5)$, with the primary eclipse centered at $\rho_\emptyset$. In addition to meeting the efficiency requirements, the phase binning process also minimizes $\rho_\emptyset$ error by smoothing over uncorrected systematics that may present a stray minimum flux below the true primary eclipse depth. 

The PGM therefore fits the phase binned curves to piecewise smooth $2^{nd}$ order polynomial chains using the \texttt{polyfit} algorithm of \citet{Prsa2008}. However, the ANN's use of the sigmoid activation function (see Sec.~\ref{ann}) leads to errors in processing large polynomial coefficients that may arise during polynomial fitting. Hence, we modify the \texttt{polyfit} algorithm to fit polynomial chains $\mathbf{X_j}$ with components $x_i$ described by three points in a $2-$dimensional \textit{Euclidean} space (phase, flux). These three data points represent the chain's two end-points, or knots, and the midpoint. Here each chain shares one knot with the next, thus a single chain is effectively described by two points in a plane. We add the additional modification of allowing either a four-polynomial chain $\{\mathbf{X_0},\mathbf{X_1},\mathbf{X_2},\mathbf{X_3}\}$ or two-polynomial chain $\{\mathbf{X_0},\mathbf{X_1}\}$ fit based on a $\chi^2$ goodness of fit criterion. With these modifications the PGM processes phased light curves initially containing on order $10^3\sim10^4$ data points into at most four linearly separable patterns $\mathbf{X_j}$ where $N=4$ for each $j = 1,2,3,4$. That is, we reduce the recognizable pattern of the phase folded light curve which may contain a very large number of data points to a maximum of sixteen points. 
 
Bin size selection is an adjustable parameter. This option allows for a condensed light curve based on the survey cadence and/or duration in order to manage the point density per bin. Of course, more bins lead to longer run times for the PGM and fewer bins lead to sparse light curves where the narrow eclipses of wide EBs might be smoothed over or missed. The minimum number of points per bin may also be adjusted in an attempt to capture narrow eclipses, however from trial and error we find that no fewer than 50 phase bins and no fewer than three bins per chain is optimal. 

\begin{figure}
\plottwo{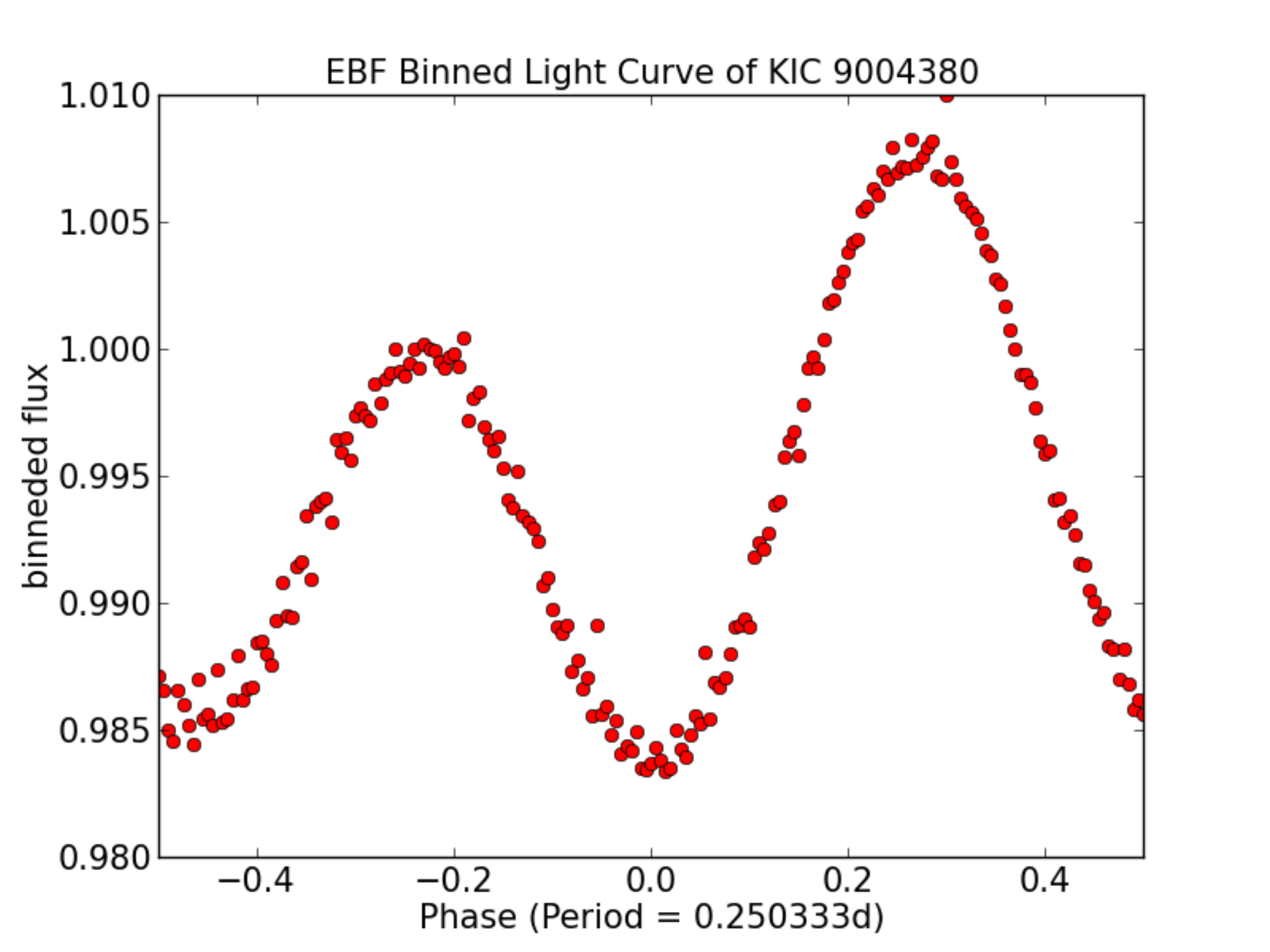}{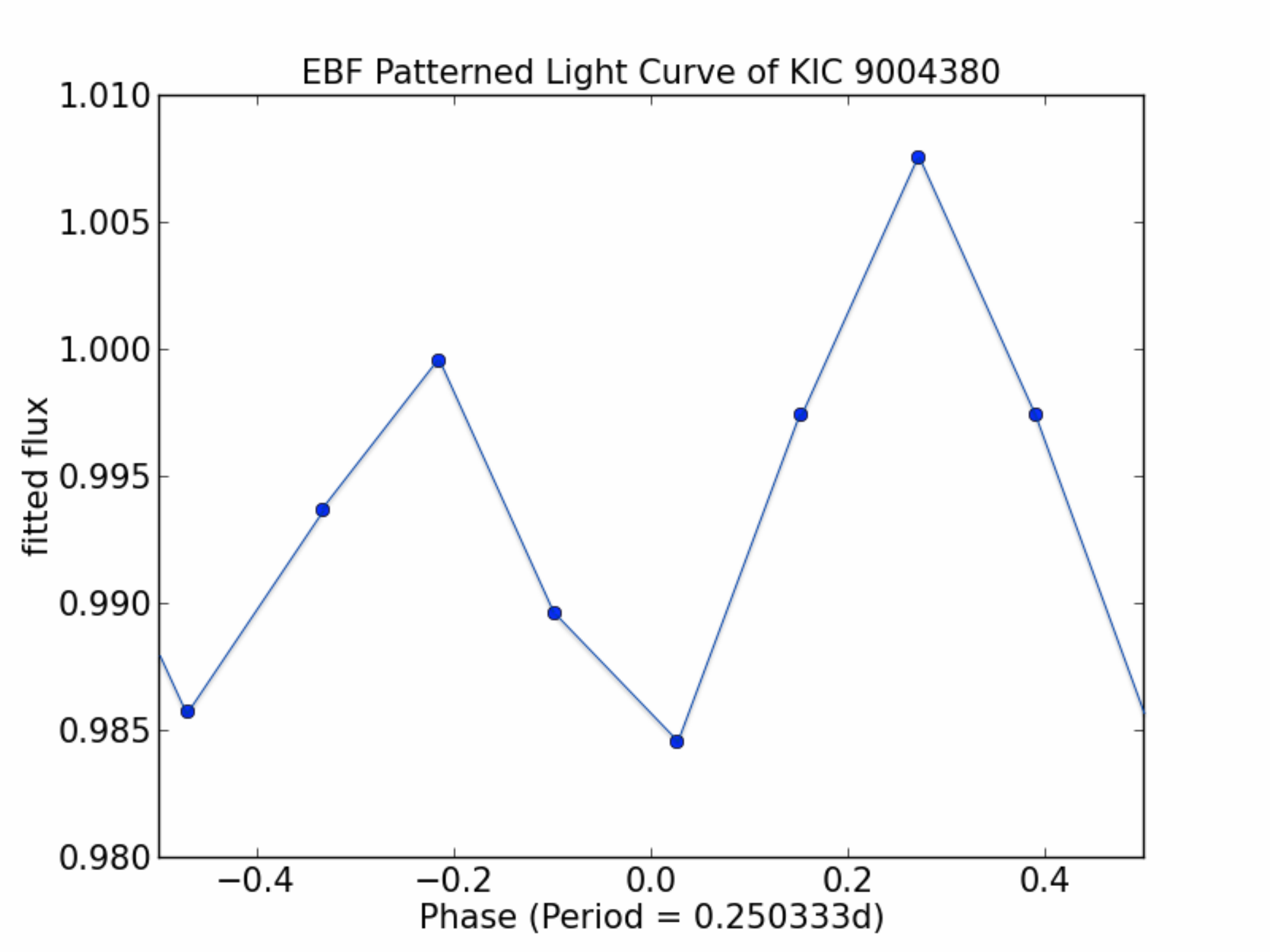}
\caption{\footnotesize{\textbf{Left:} The PGM generated binned flux light curve for KIC 9004380 with 200, size $= 0.005$, phase bins. \textbf{Right:} The PGM generated linearly separable patterns, plotted as the knots and midpoints for KIC 9004380 with a minimum of three phase bins per polynomial chain.}\label{Figure5}}
\plottwo{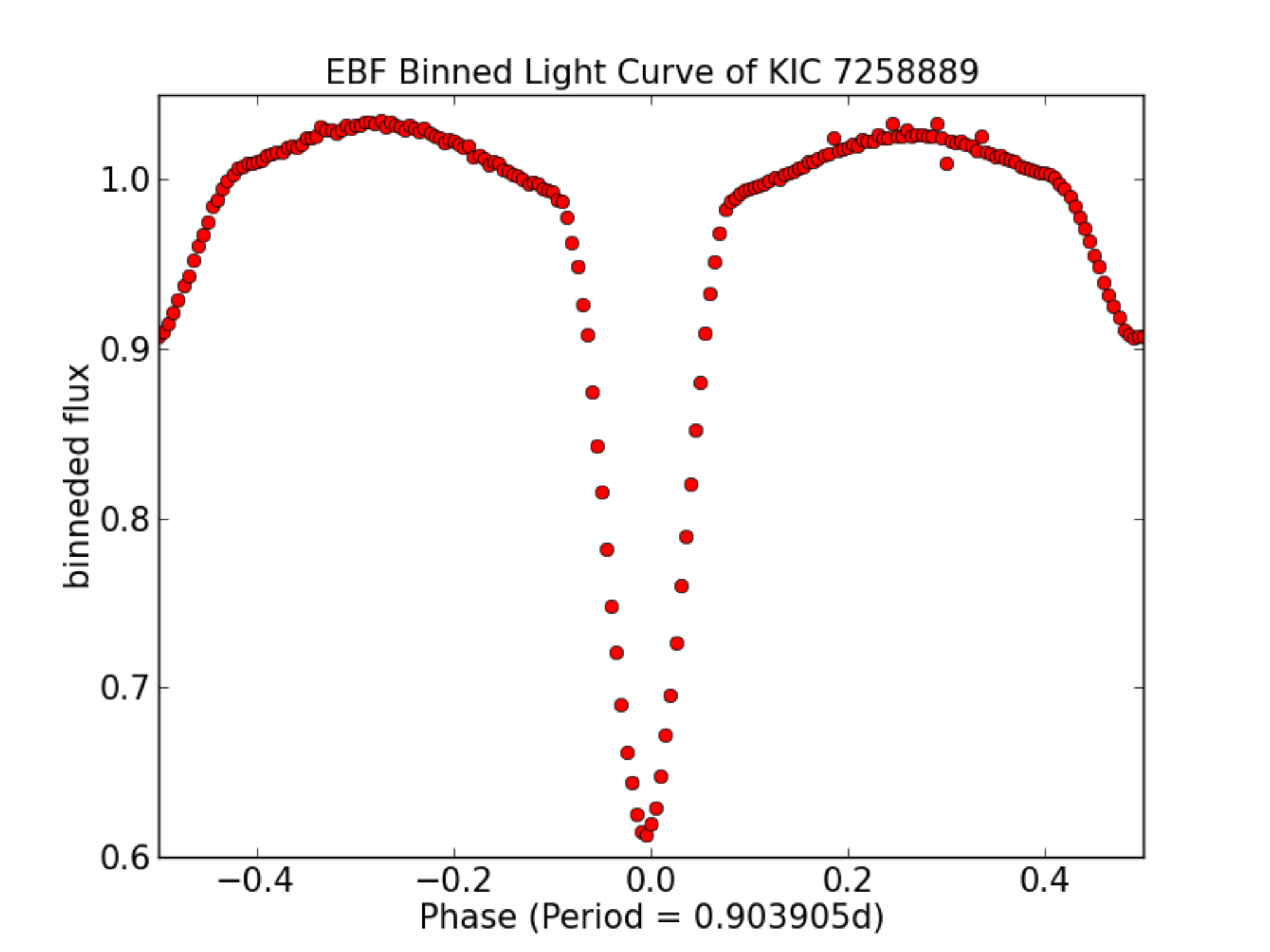}{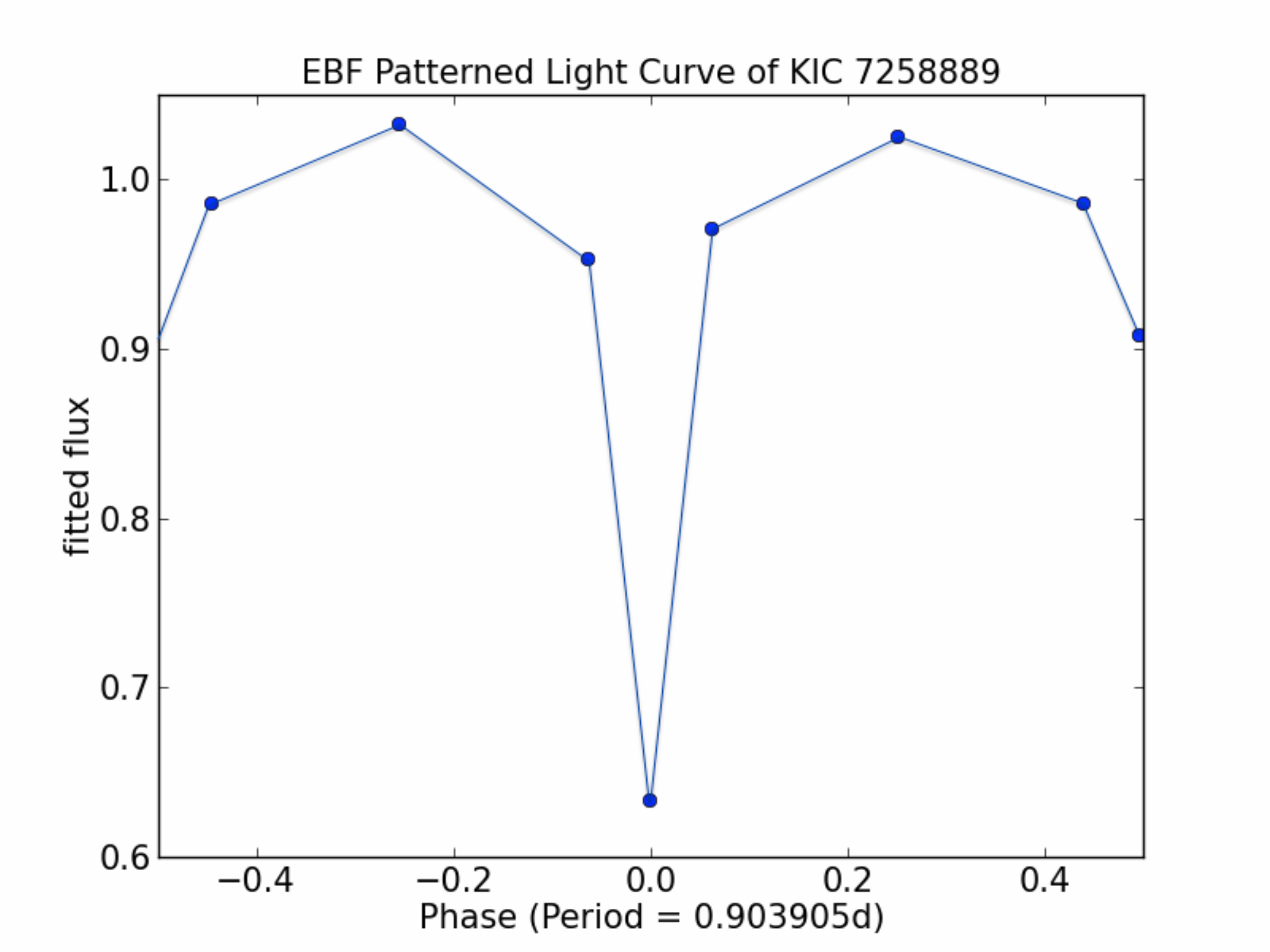}
\caption{\footnotesize{\textbf{Left:} The PGM generated binned flux light curve for KIC 7258889 with 200, size $= 0.005$, phase bins. \textbf{Right:} The PGM generated linearly separable patterns, plotted as the knots and midpoints for KIC 7258889 with a minimum of three phase bins per polynomial chain.}\label{Figure6}}
\plottwo{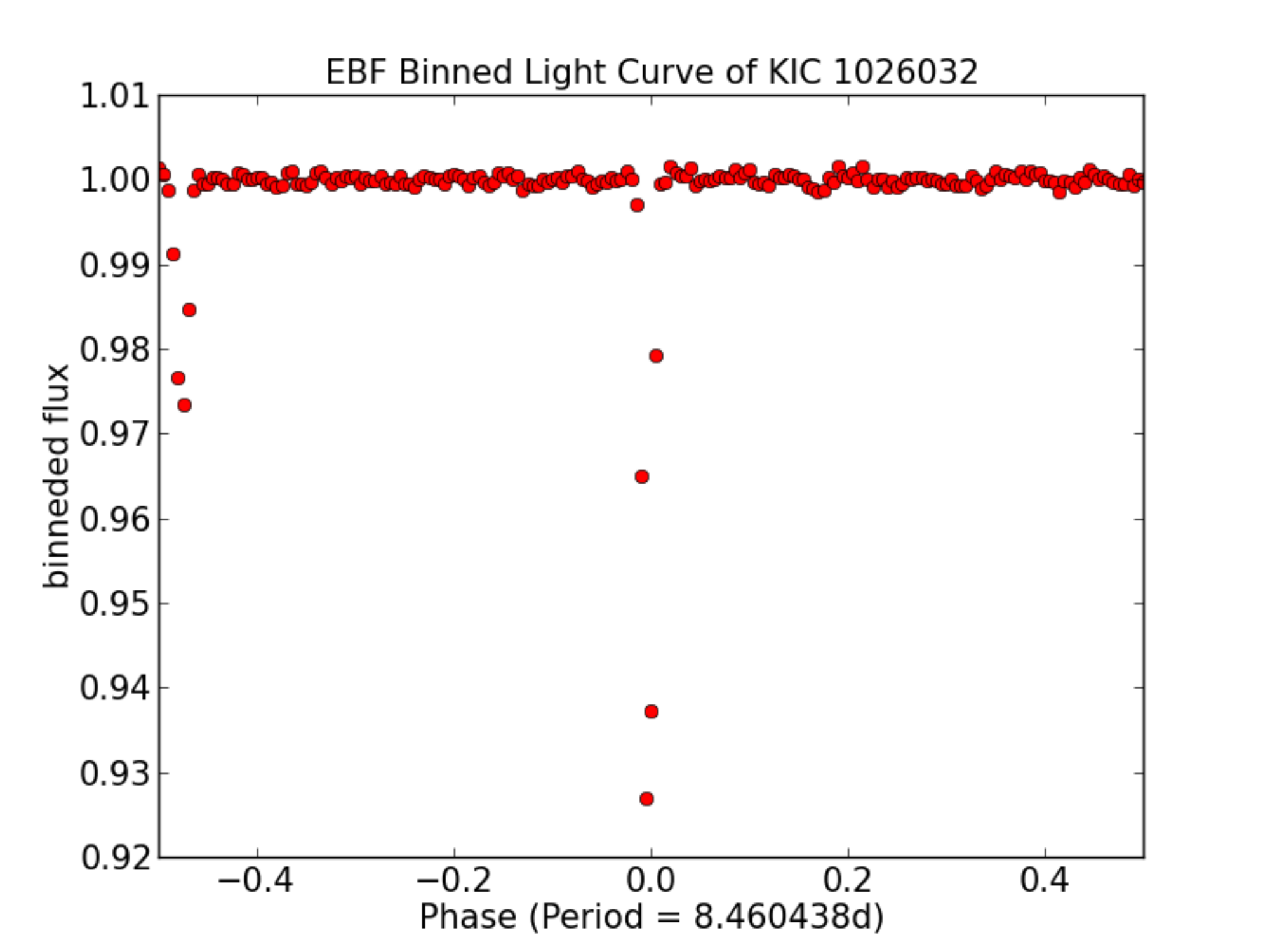}{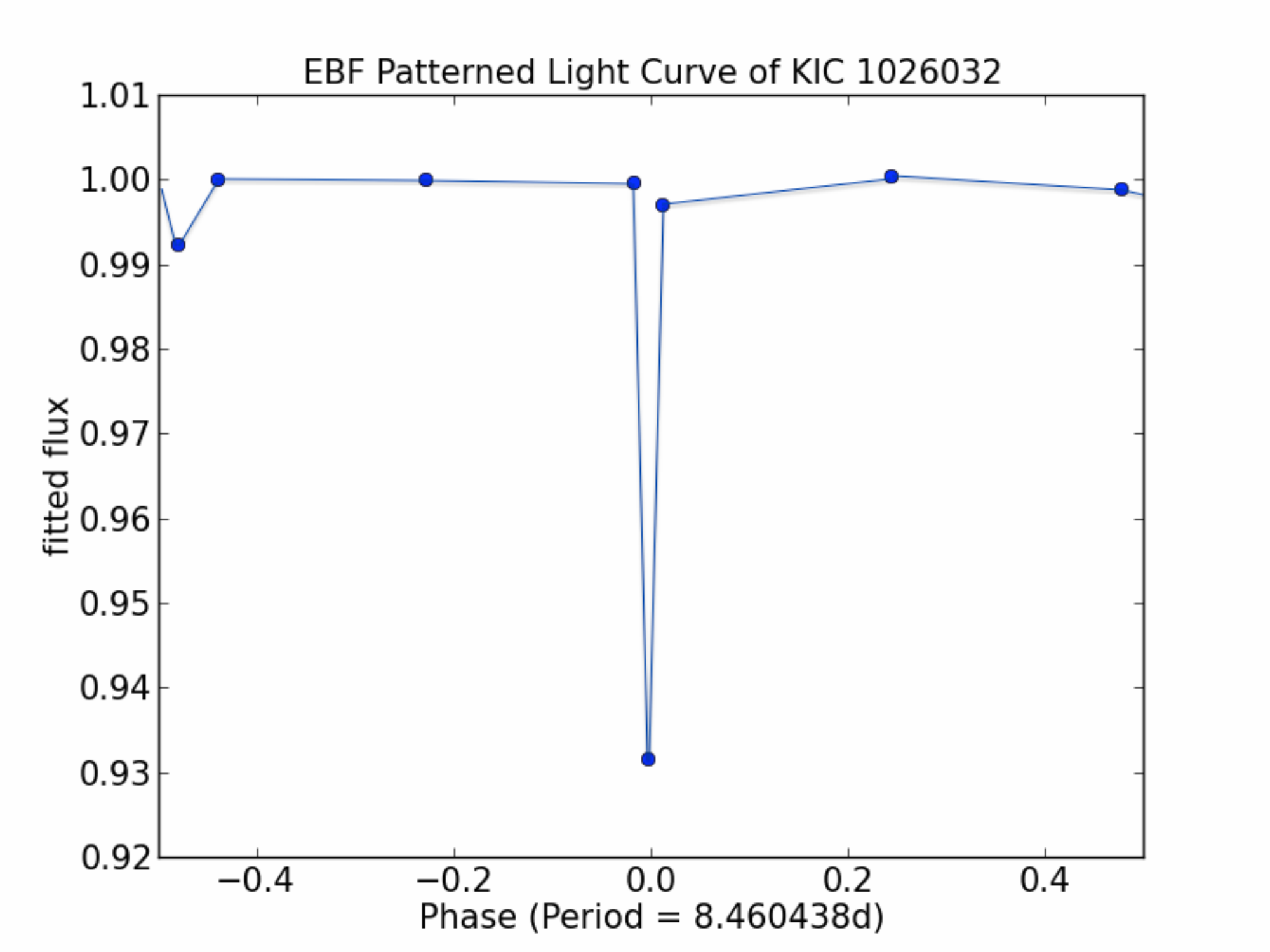}
\caption{\footnotesize{ \textbf{Left:} The PGM generated binned flux light curve for KIC 1026032 with 200, size $= 0.005$, phase bins. \textbf{Right:} The PGM generated linearly separable patterns, plotted as the knots and midpoints for KIC 1026032 with a minimum of three phase bins per polynomial chain.}\label{Figure7}}
\end{figure}

To fit the \textit{Kepler} ``Q3'' phased light curves we bin each light curve's $\sim$5,000 long cadence data points into 200 (size $= 0.005$) phase bins with a polynomial chain that contains no fewer than three data points. The EBF generated binned light curves and linearly separable patterns, plotted as the knots and midpoints of the module's fitted flux, as examples of \textit{Kepler} EBs with the morphology EC, ESD, ED are shown in Figure \ref{Figure5}, Figure \ref{Figure6}, and Figure \ref{Figure7}, respectively.

\begin{figure}[!ht]
\begin{center}
\includegraphics[scale=0.7]{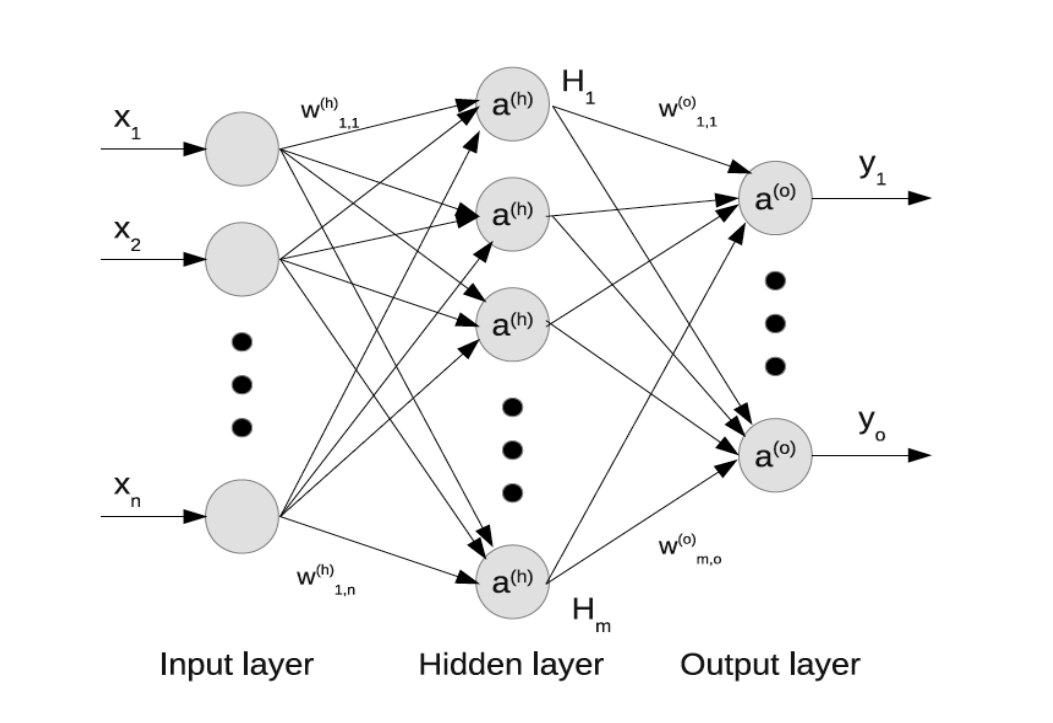}
\caption{\footnotesize{The EBF single layer perceptron neural network classifier with input vector $\mathbf{X}$, ridge function derived feature $\mathbf{H}$, and output $\mathbf{Y}$; where $n = 19$, $m = 20$, $o = 10$, and $a(h) = a(o) = \sigma (\nu) = (1 + e^{-\nu})^{-1}$.}}
\label{Figure8}
\end{center}
\end{figure}

\subsection{Artificial Neural Network Classifier (Pattern Recognition Module) \label{ann}}

The EBF's trained, validated, and tested ANN is the single hidden layer feed forward perceptron network diagrammed in Fig. \ref{Figure8}. This module takes as input the PGM output set of data points as a simple representation of the target light curve. These patterned inputs are recognized by the ANN and classified into the EB categories that it has been trained to recognize. The outputs of the ANN are therefore the 
classification of each light curve with an associated confidence level in the classification. A detailed description of the ANN,
including its training, validation, and output classifications, is described in \citet{Paegert2014}. Here we provide a summary of its specifications and functions.

The ANN functions as a two-stage pattern recognition classifier based on nonlinear statistical models. The input layer, nodes $x_1$ to $x_n$, are populated with the PGM fitted polynomial chains $\mathbf{X_j}$ as an input vector of $n$ components that is then propagated via weighted connections $w(h)$ to the hidden layer. The activation function $a(h)$ is in the form of the sigmoid 
\begin{equation}
\sigma (\nu) = (1 + e^{-\nu})^{-1}
\end{equation}
where $\nu$ is the variable used by the ridge function to derive the features $H_1$ to $H_m$ as linear combinations of the inputs. The signals are again weighted by $w(o)$ and propagated to the output layer where there are $o$ target classifications coded as a 0 to 1 categorical target variable $\nu$ for the $o^{th}-$class classification using the same sigmoid activation function to model the target nodes $y_1$ to $y_o$ as linear combinations of $H_1$ to $H_m$; where $n = 19$, $m = 20$, and $o = 10$. In other words, the ANN accepts $n=19$ input parameters --- representing the 16 polynomial chain parameters from the PGM as well as the logarithm of the period in days, the total $\chi^2$ as a measure of fit, and the maximal amplitude of the light curve's normalized flux --- and outputs a probability associated with each of the $o=10$ known light curve classifications. The 10 known classifications are the three EB categories (EC, ESD, ED), the other variable types (DCEP-FU, DCEP-FO, DSCT, RRAB, RRC, MIRA), and a miscellaneous classification (MISC).

Training of the ANN was conducted by back-propagation, and gradient descent was used to minimize the error in the set of all weights $\theta$, or $err_{\theta}$, where the error function is given by
\begin{equation}
err_{\theta} = R(x,y) = 	\sum\limits_{i=1}^n\sum\limits_{j=1}^o(y_{ij} - f(x_i)_j)^2   
\end{equation}
for $i = 1,2,3,\ldots n$ and $j = 1,2,3,\ldots o$. $f(x_i)_j$ is described by the softmax function 
\begin{equation}
f(x_i)_j  = g(T)_k = \frac{e^{T_k}}{\sum\limits_{k=1}^o e^{T_k}}  
\end{equation}
as $T \rightarrow \infty$, and was used to approximate the nonlinear input functions. As described in \citet{Paegert2014}, training and validation of the ANN was done using 32,278 light curves from the Automated All-Sky Survey (ASAS) \citep{Pojmanski1997}. These tests showed the ANN is capable of an accuracy of 90.6\%, 76.5\%, and 90.3\%, and a completeness of 95.8\%, 63.7\%, and 87.8\% for ECs, ESDs, and EDs, respectively.

In summary, for each input patterned light curve from the PGM, the ANN outputs a most likely classification and an associated confidence in that classification. The output classifications for EBs can be either EC, ESD, or ED. Because the ANN was originally developed on the ASAS sample \citep{Paegert2014}, there is also a classification of MISC for light curves that are not clearly identified as one of the three EB classes. In addition, the ANN outputs up to 9 additional possible classifications (and associated confidence levels) for each light curve. As we describe below, this permits us to identify possible EBs at lower confidence that, e.g., have a primary classification of MISC but secondary classification of EC, ESD, or ED.

\subsection{Candidate Validation Module \label{cvm}}

Once ANN classifications are complete, candidate EC, ESD, and ED light curves are filtered for false positives by the candidate validation module (CVM). We position the CVM as a post classification process to provide the option to examine some or all of the filtered candidates. 

The CVM filters candidates by two distinct metrics. The first is the \texttt{polyfit} algorithm's total goodness of fit, $\chi_T^2$, such that 
\begin{equation}
\chi_T^2 \equiv \sum\limits_{j=0}^{m} \sum\limits_{i=1}^{l} w_{ji}(x_{ji}-\Phi_{ji})^2;
\end{equation}
where $\Phi_{ji}$ is the normalized, detrended flux of phase bin $l$ and $x_{ji}$ is the $i^{th}$ component of the $j^{th}$ polynomial chain $\mathbf{X_j}$ for $j=0,1..,m$ that satisfies the requirement of linear separability from Section \ref{pgm}. The $\chi_T^2$ metric describes how well the PGM represents the candidate light curves and is a useful measure of whether the ANN's recognition of the EBF generated pattern translates to an accurate classification as an EB. 

The second metric is the normalized, detrended light curve's scaled sum of all second derivatives for the portion of the curve between each and every data point, which we refer to as $\delta^2$ and defined as 
\begin{equation}
\delta^2 \equiv \frac{1}{\Phi_{max} - \Phi_{min}} \sum\limits_{i=2}^{n-1} \Bigg\vert \frac{1}{\rho_{i+1}-\rho_{i-1}} \bigg(\frac{\Phi_i-\Phi_{i-1}} {\rho_i-\rho_{i-1}} - \frac{\Phi_{i+1}-\Phi_i}{\rho_{i+1}-\rho_i}\bigg) \Bigg\vert
\end{equation}
where $\rho_i$ and $\Phi_i$ are the phase and flux for the $i^{th}$ of $n$ data points. As such, $\delta^2$ is a measure of the dispersion in the detrended light curve and when compared to the average detrended flux measurement uncertainty calculated in the DCM provides a criterion for filtering a pseudo-periodic signal that may arise from random noise. 

\begin{figure}[!hb]
\plottwo{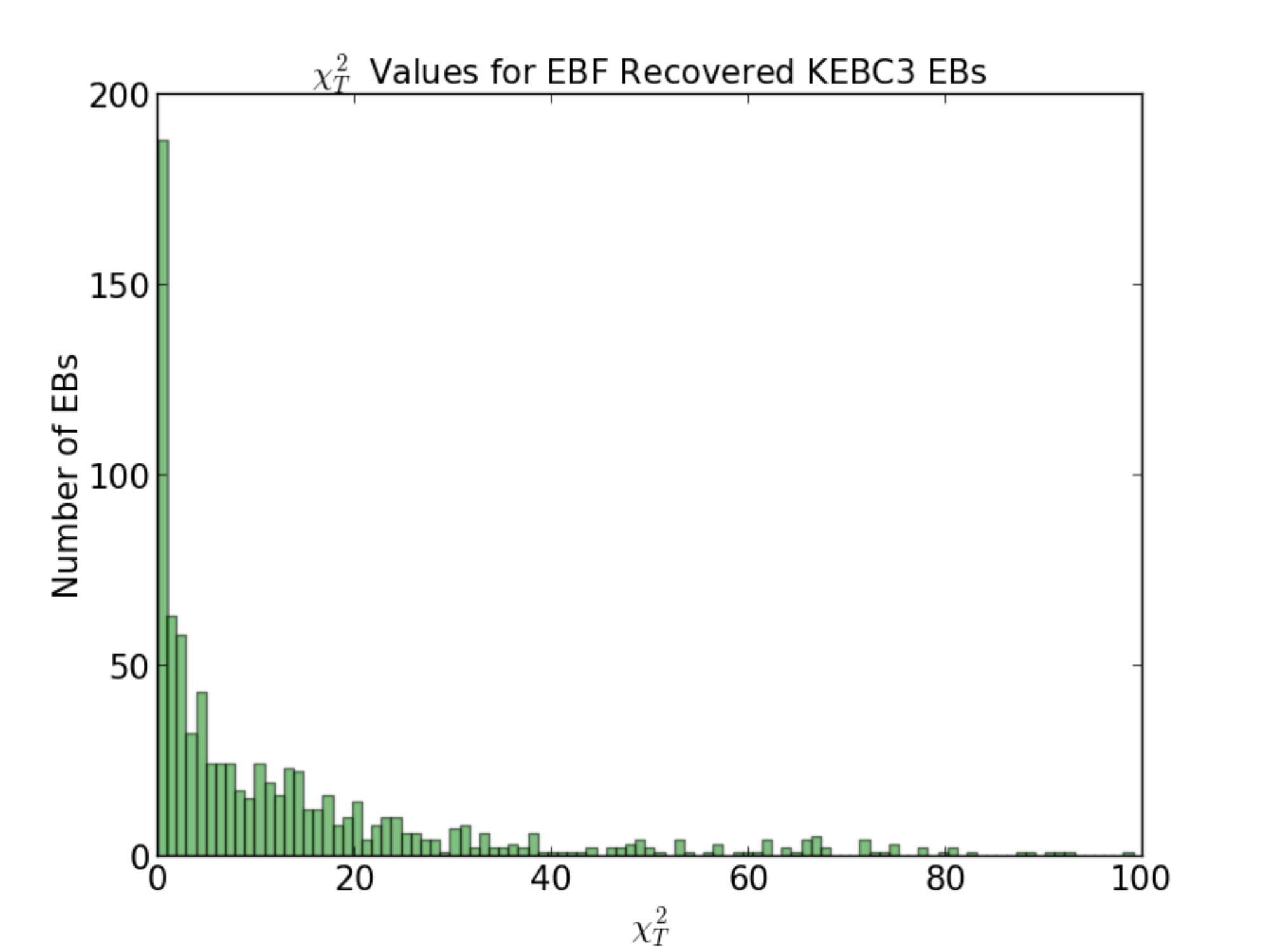}{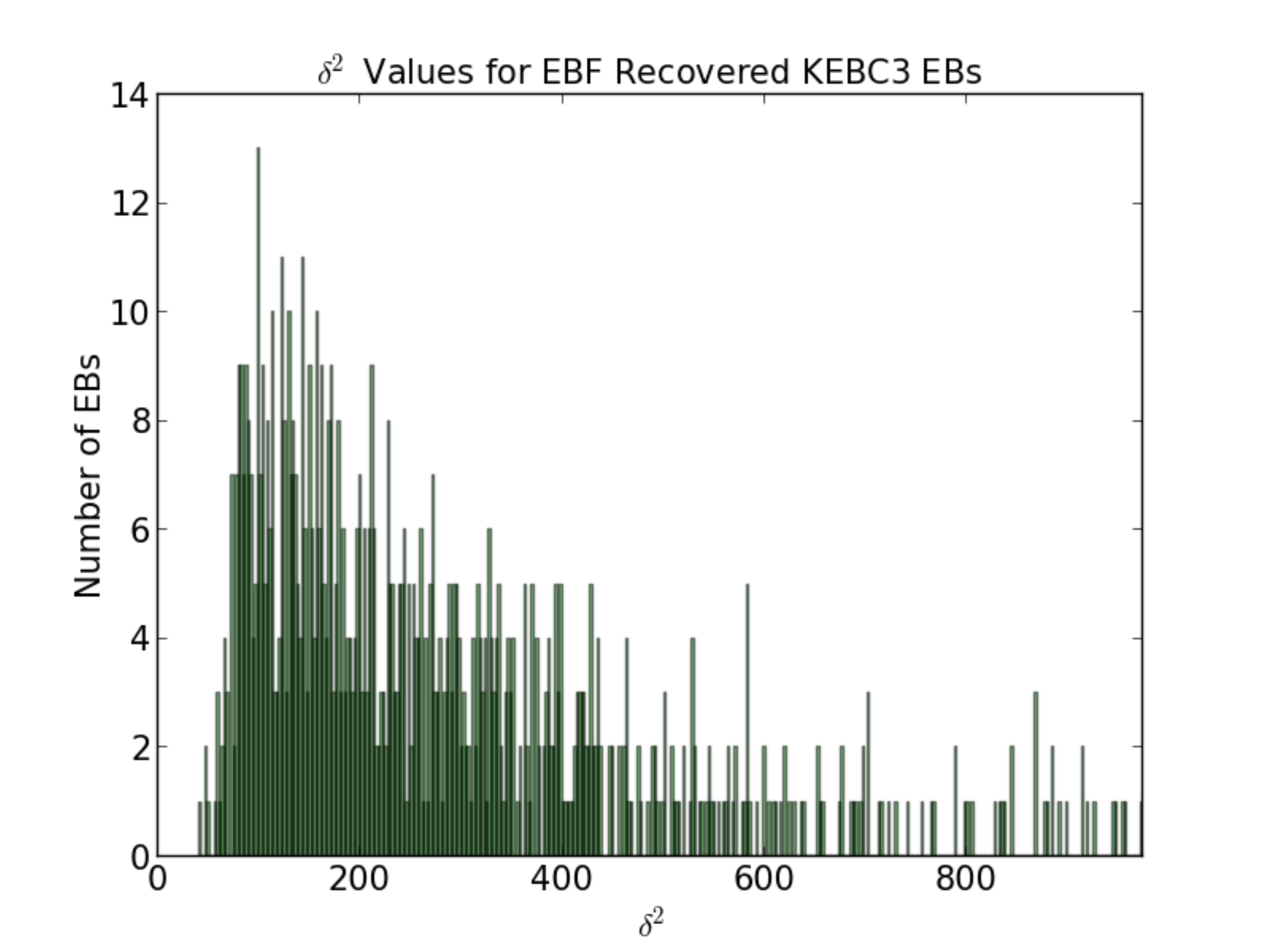}
\caption{\footnotesize{\textbf{Left:} Histogram of the PGM $\chi_T^2$ over the ANN recovered sample of KEBC3 EBs used to calculate the CVM filter criteria of $\chi_T^2 \ge 100$  \textbf{Right:} Histogram of the CVM $\delta^2$ over the ANN recovered sample of KEBC3 EBs used to calculate the CVM filter criteria of $\delta^2 \ge 1,000$.}\label{Figure9}}
\plottwo{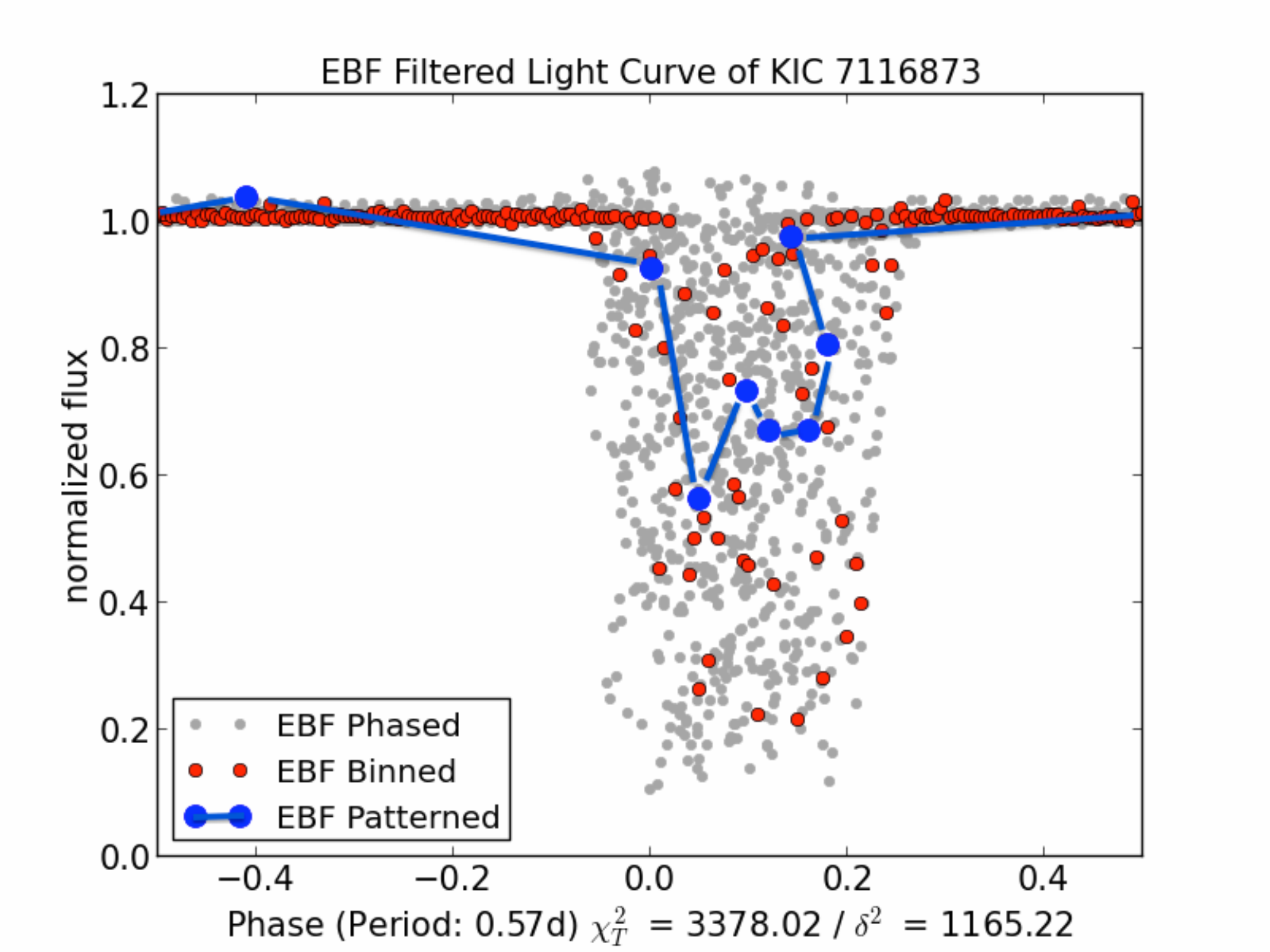}{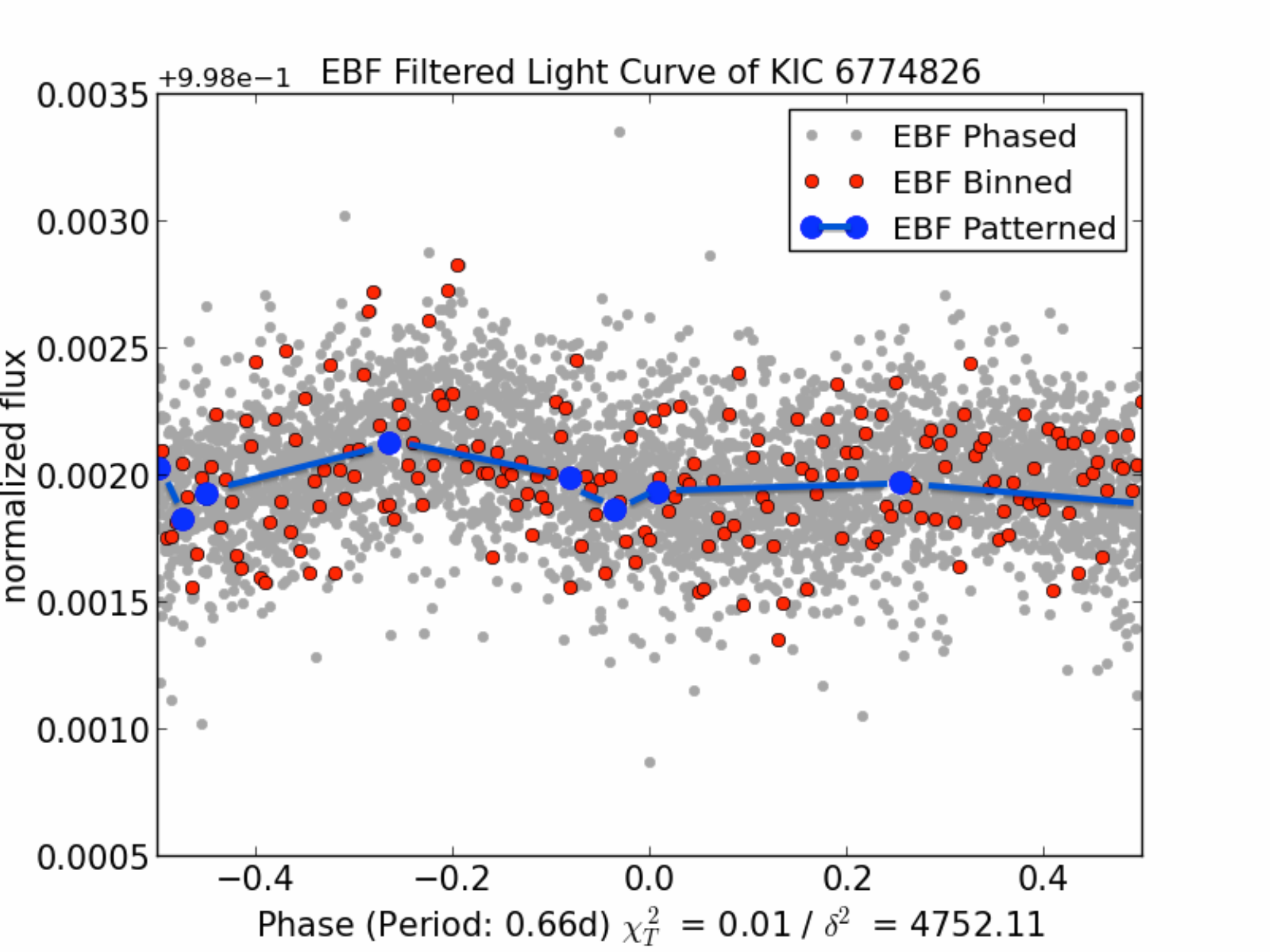}
\caption{\footnotesize{\textbf{Left:} The EBF generated light curve of KIC 7116873 where the PGM fits to some quasi-systematic artifact that the ANN appropriately classifies as an ESD, however the CVM criteria of $\chi_T^2 \ge 100$ filters this candidate as a false positive.   \textbf{Right:} The EBF generated light curve of KIC 6774862 where the PGM fits to a pseudo-sinusoidal signal that the ANN appropriately classifies as an EC, however the CVM filter criteria of $\delta^2 \ge 1,200$ prevents this false positive from populating the automated EB catalog.}\label{Figure10}}
\end{figure}

The CVM adapts the $\chi_T^2$ and $\delta^2$ filter metrics by using $\sim30$ known EBs contained in the survey. The module then calculates the upper limits on both the $\chi_T^2$ and $\delta^2$ from this benchmark subset. A CVM analysis of the KEBC3 EBs contained in the \textit{Kepler} ``Q3'' data set and recovered by the ANN provides a $\chi_T^2$ upper limit of 100 and a $\delta^2$ upper limit of 1,200 as shown by the histograms in Figure \ref{Figure9}. 

Figure \ref{Figure10} shows the light curves of two example \textit{Kepler} targets where the PGM fits to some quasi-systematic artifact that the ANN consequently mis-classifies as an ESD and where the PGM fits to a pseudo-sinusoidal signal generated by random noise that the ANN consequently mis-classifies as an EC. These examples demonstrate the CVM determined criteria of $\chi_T^2 \ge 100$ and $\delta^2 \ge 1,200$ filtering false positives from the automated EB catalog. Notice that a CVM filter with the criteria of only $\chi_T^2 \ge 100$ would have passed the target, with $\chi_T^2$ of only 0.01, yet a $\delta^2$ of 4,752 to the EBF catalog. Similarly, a CVM criteria of only $\delta^2 \ge 1,200$ would have passed the target, with a $\delta^2$ of 1,165 yet a, $\chi_T^2$ of 3,378 to the EBF catalog. Thus, we set the CVM cut-off parameters to $\chi_T^2 = 100$ and $\delta^2 = 1,200$ in order to minimize EB classification false positives.

\section{Results \label{RES}}

We use the 2,612 EBs cataloged in the KEBC3 to benchmark the EBF pipeline outputs as an initial assessment of pipeline capabilities on standard morphology ECs, ESDs, and EDs. Thus, we first remove from the benchmark sample those KEBC3 EBs flagged as eclipse timing variations (ETV) induced by third body candidates \citep{Conroy2013}, heart beat (HB), and those classified in the KEBC3 as uncertain (UNC). We also exclude those KEBC3 EBs with cataloged periods outside the \textit{Kepler} ``Q3'' duration parameters of $0.11d\le$ period $\le20.1d$ and of course those cataloged EBs that do not appear in the \textit{Kepler} ``Q3'' data set. Thus, we benchmark the EBF pipeline against an adjusted KEBC3 sample of 1,198 cataloged EBs. 

\subsection{EBF Performance Metrics}

\subsubsection{Pre-Classification Metrics from the DCM, HRM, and PGM \label{resp}}

We compare the metrics of the EBF pipeline's pre-classification output to the same metrics of the manually determined KEBC3. Note that the KEBC3 light curves contain $\sim50,000$ data points from more than ten quarters of observations while the ``Q3'' observations include only $\sim5,000$. 

We evaluate the DCM's performance by comparing the 200 phase binned, normalized flux values of each common target. We find the DCM's performance to be in excellent agreement, mean $\chi^2 = 0.00212$, with the KEBC3 benchmark. This DCM light curve agreement breaks down to a mean $\chi^2$ of 0.00373, 0.00264, 0.00052 for cataloged ECs, ESDs, and EDs respectively.

\begin{figure}[!ht]
\plottwo{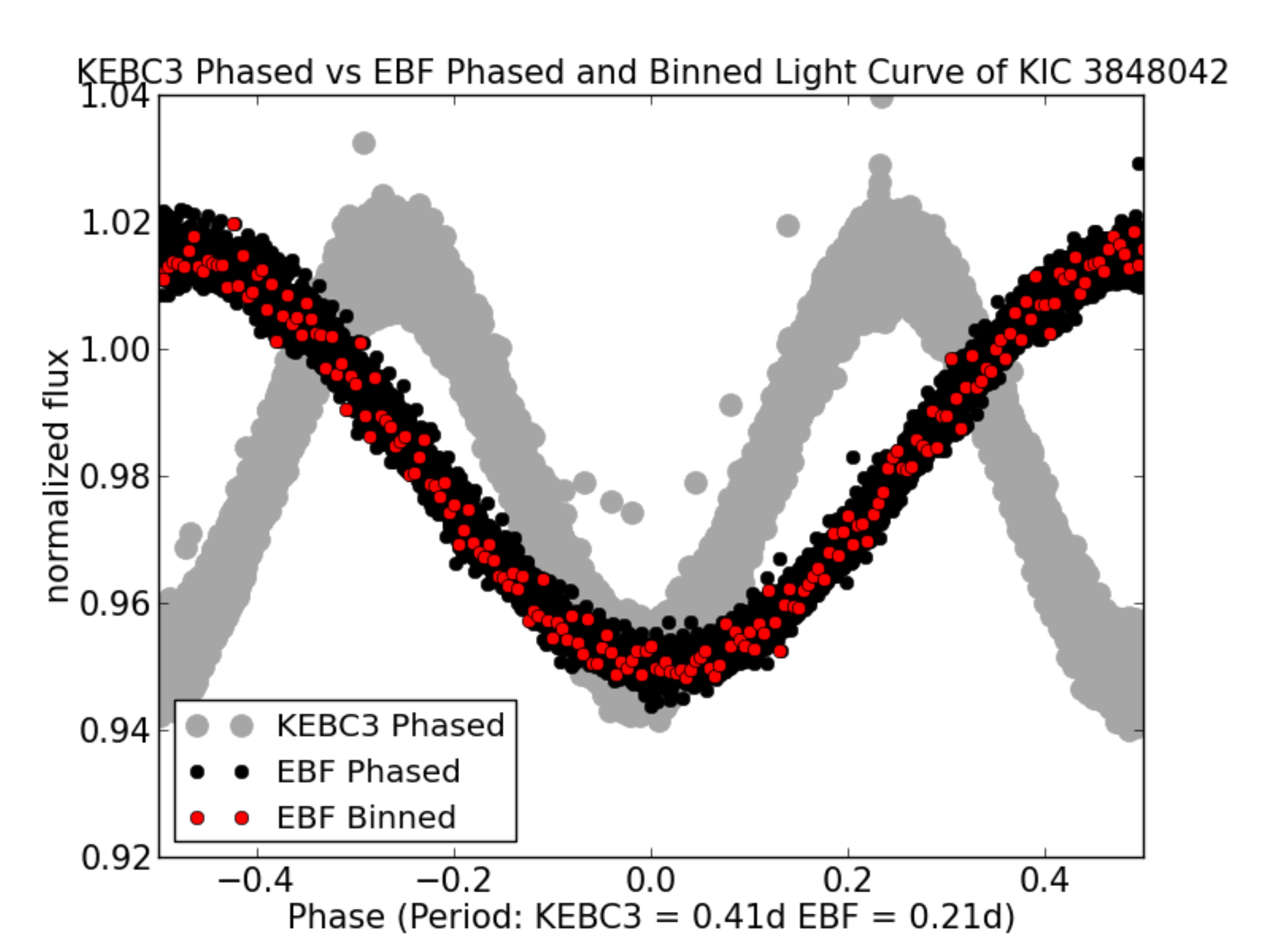}{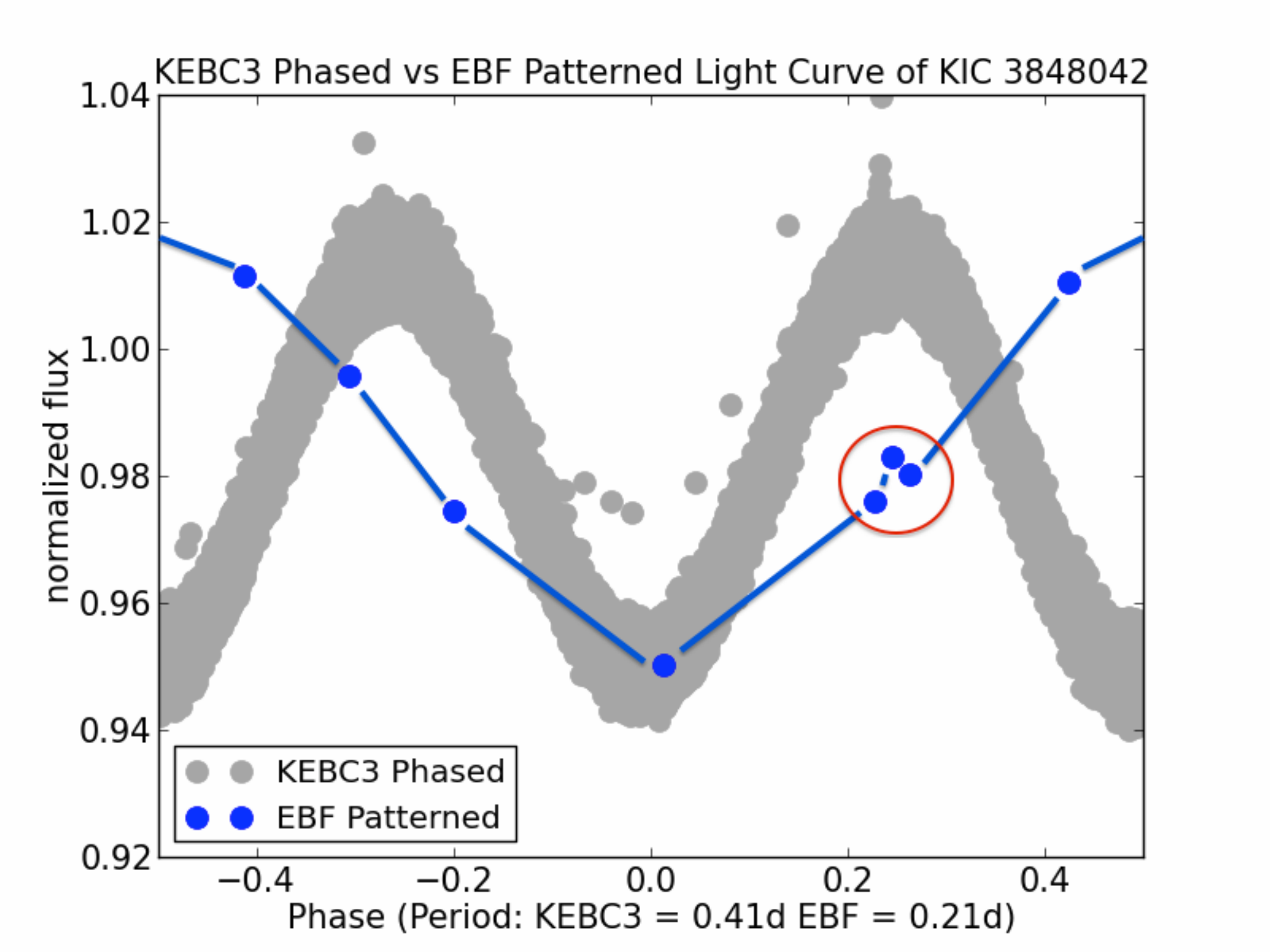}
\caption{\footnotesize{\textbf{Left:} KEBC3 light curve vs. EBF phased, binned light curve of KIC 3848042 detrended by sigma-clipping over $(-1\sigma, 3\sigma)$ interval of the \textit{Legendre} function $P_{10}(x)$ with 200 phase bins. \textbf{Right:} KEBC3 light curve vs. EBF patterned curve of KIC 3848042 fit to polynomial chains of no less than three bins. Here the EBF recovers the cataloged EC in spite of a half-period alias and a small aberrant $3^{rd}$ chain circled in red.}\label{Figure11}}
\end{figure}

\begin{figure}[!ht]
\plottwo{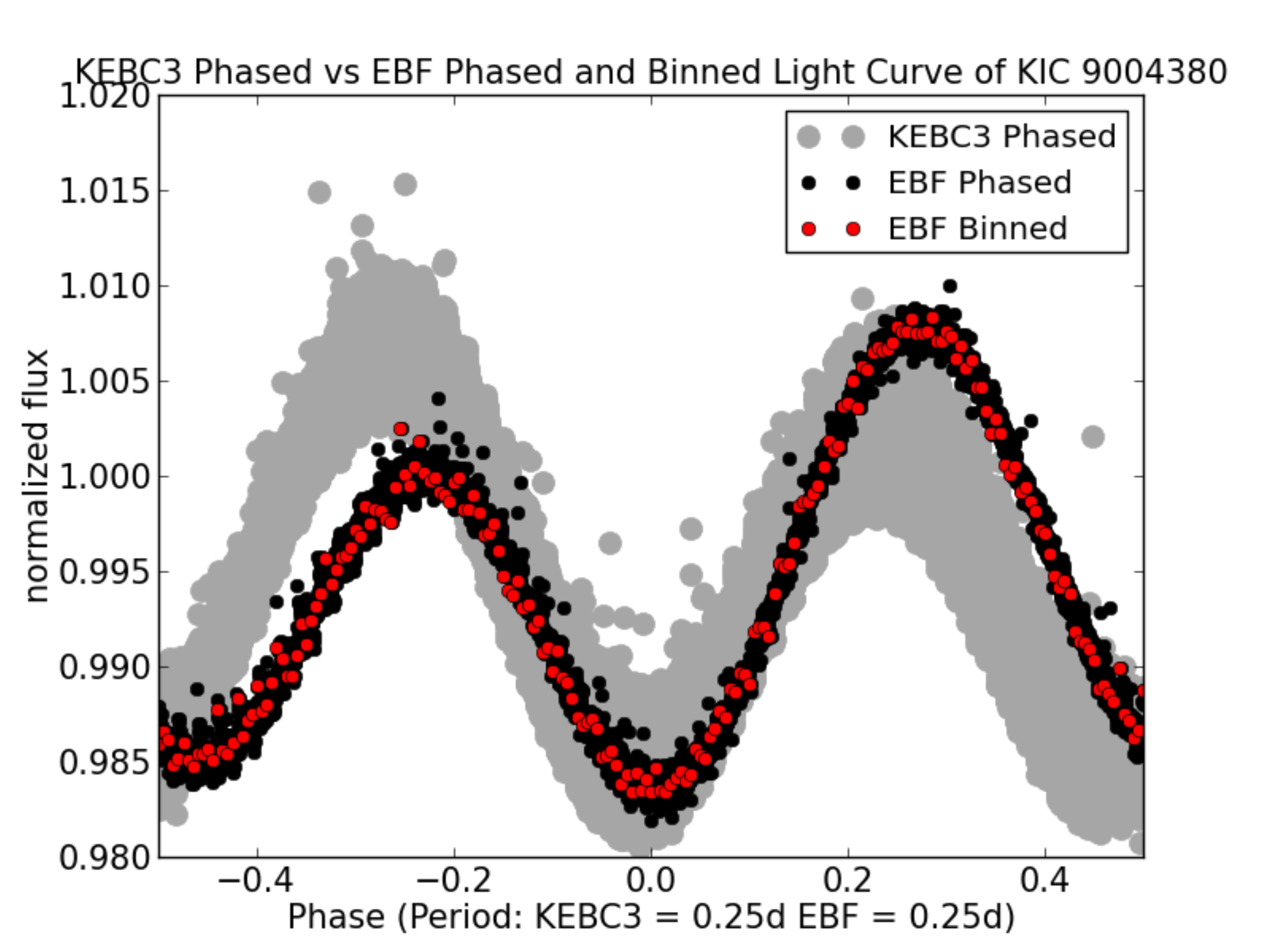}{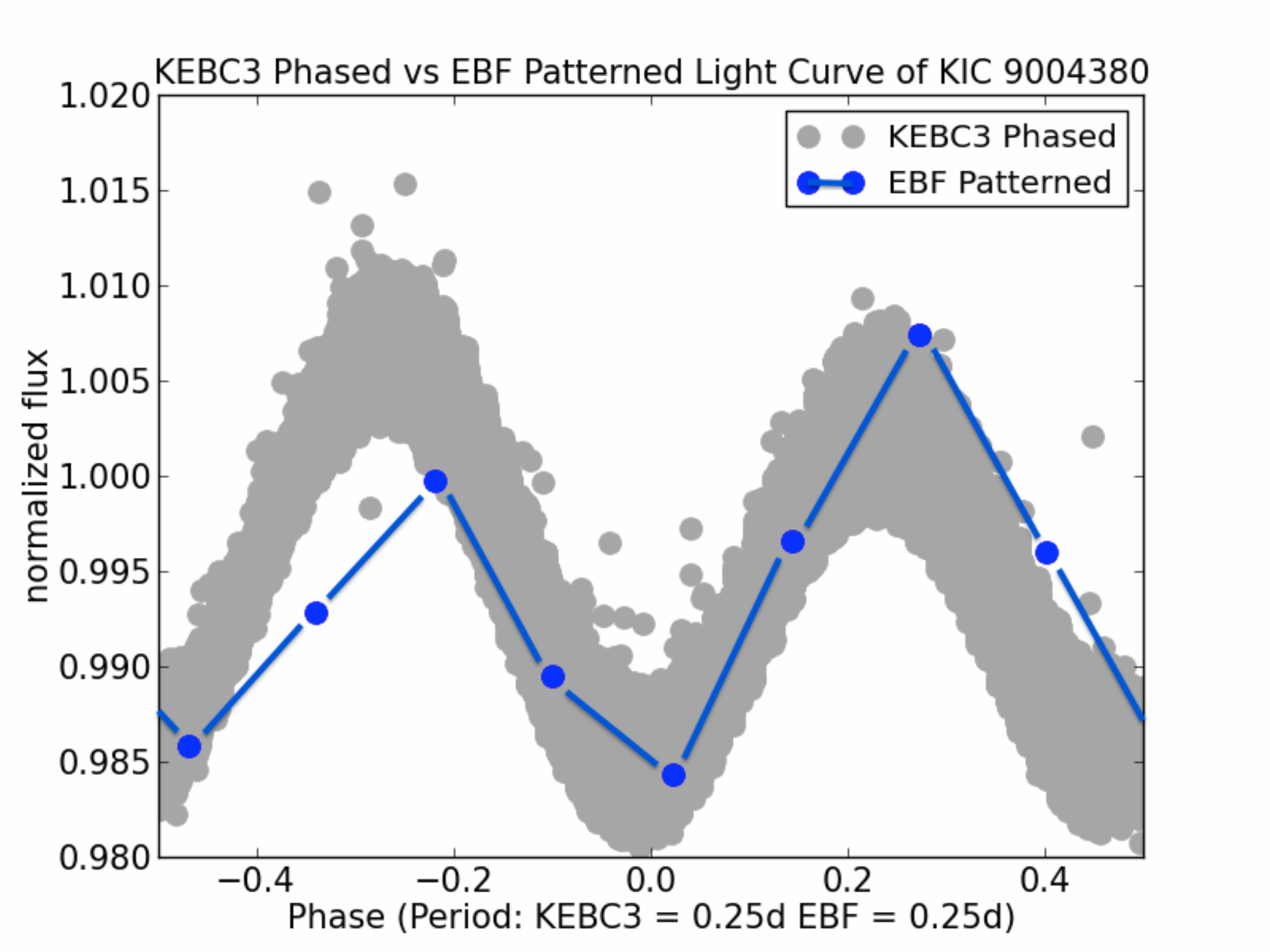}
\caption{\footnotesize{\textbf{Left:} KEBC3 light curve vs. EBF phased, binned light curve of KIC 9004380 detrended by sigma-clipping over $(-1\sigma, 3\sigma)$ interval of the \textit{Legendre} function $P_{10}(x)$ with 200 phase bins. \textbf{Right:} KEBC3 light curve vs. EBF patterned curve of KIC 9004380 fit to polynomial chains of no less than three bins. Here the EBF recovers the cataloged EC in spite of  a $\rho_\emptyset$ disagreement with the manually adjusted benchmark.}\label{Figure12}}
\end{figure}

The HRM assigned period agrees to within $\pm0.09d$ for 99.67\% of the EBF identified periodic variable candidates found in the KEBC3, though the HRM failed to identify statistically significant periodicity in 5.59\% of the benchmark sample, and therefore were not considered for follow on classification by the ANN. A further comparison of the HRM assigned periods to those manually determined in the KEBC3 reveals a period alias rate of 21.65\%, (i.e., the percentage of EBF determined periods that are an alias of the KEBC3 period). However, aliasing does not negatively impact ANN pattern recognition and the true period is easily determined by examining the candidate light curve. Figure \ref{Figure11} shows examples of both the EBF generated and the manually determined KEBC3 light curves for targets where the EBF assigns a half-period alias. It was also discovered that phase disagreement was common for targets with a $FR>0.5$. Though there is no adverse effect on pattern recognition for the example targeted EC in Figure \ref{Figure12}, the ambiguity in phasing a light curve with an FR of $\le0.7$ on the minimum flux led to diminished pipeline accuracy as other variable types mistaken by the pipeline for EBs (e.i., $\delta-$cepheids incorrectly phased on the flux minimum) contributed to the false positive rate.


\begin{figure}[!ht]
\plottwo{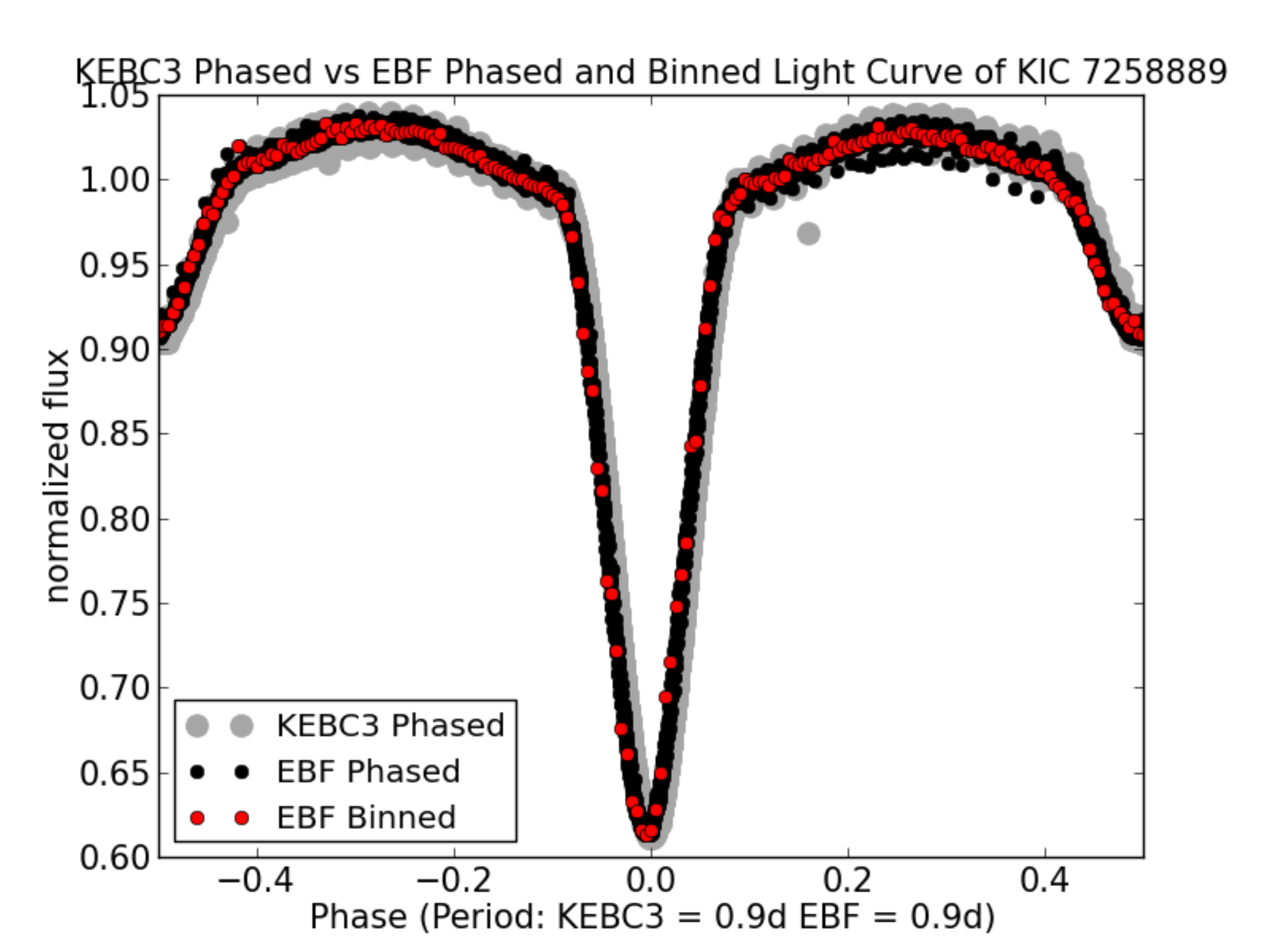}{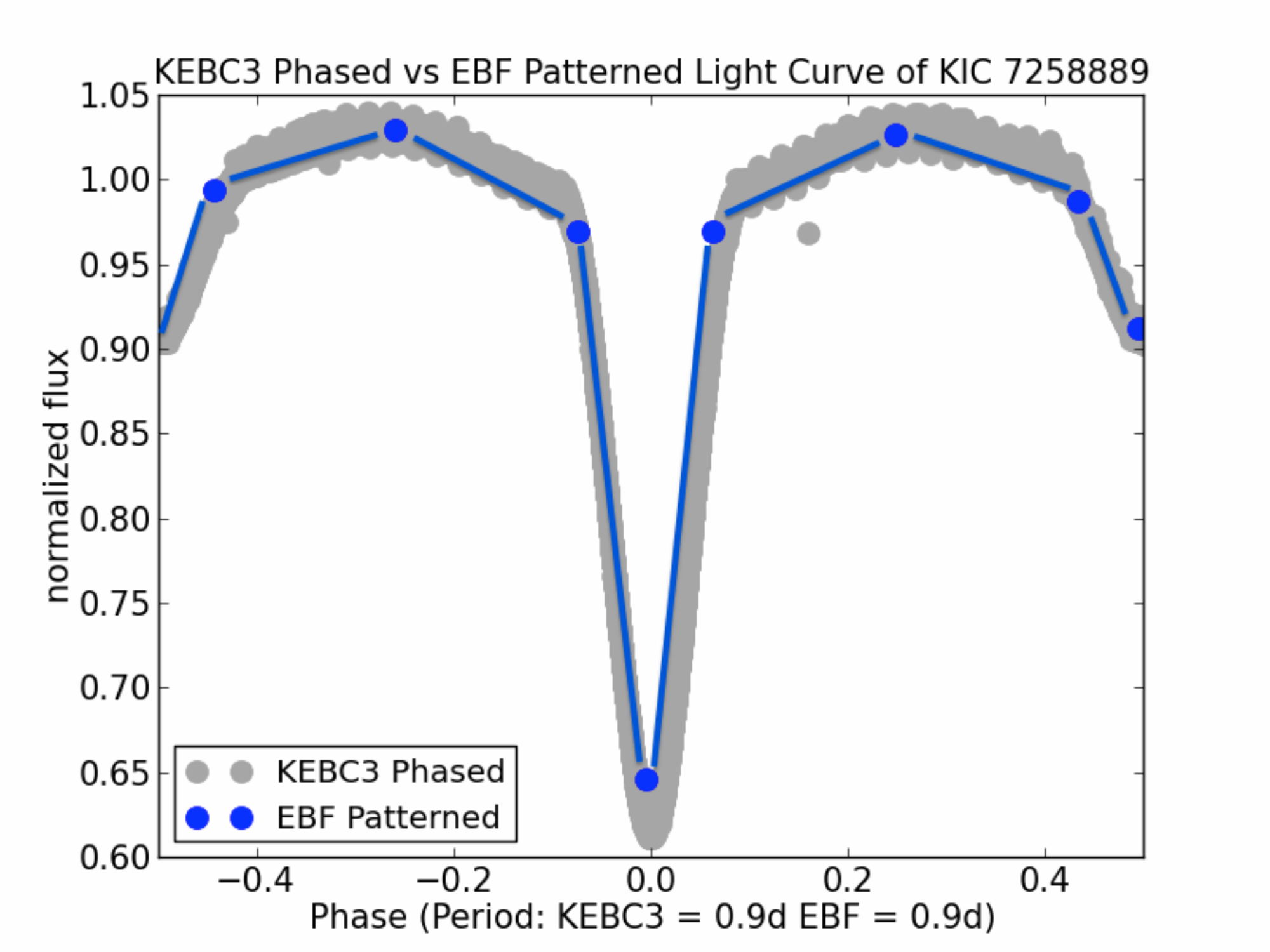}
\caption{\footnotesize{\textbf{Left:} KEBC3 light curve vs. EBF phased, binned light curve of KIC 7258889 detrended by sigma-clipping over $(-1\sigma, 3\sigma)$ interval of the \textit{Legendre} function $P_{10}(x)$ with 200 phase bins. \textbf{Right:} KEBC3 light curve vs. EBF patterned curve of KIC 9004380 fit to polynomial chains of no less than three bins. Here the EBF recovers the cataloged ESD with excellent agreement between the KEBC3 light curve and the EBF generated pattern.}\label{Figure13}}
\end{figure}

\begin{figure}[!ht]
\plottwo{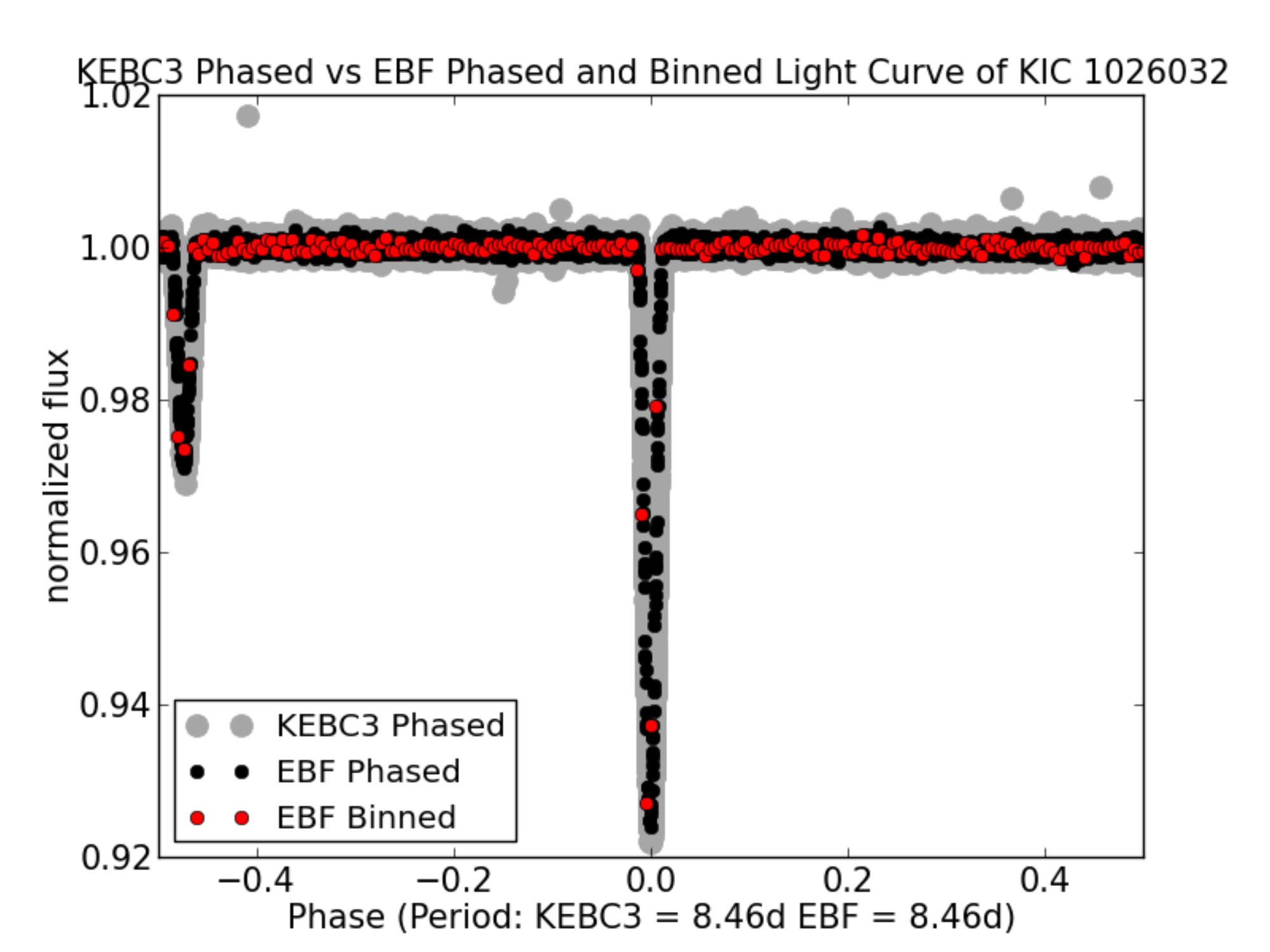}{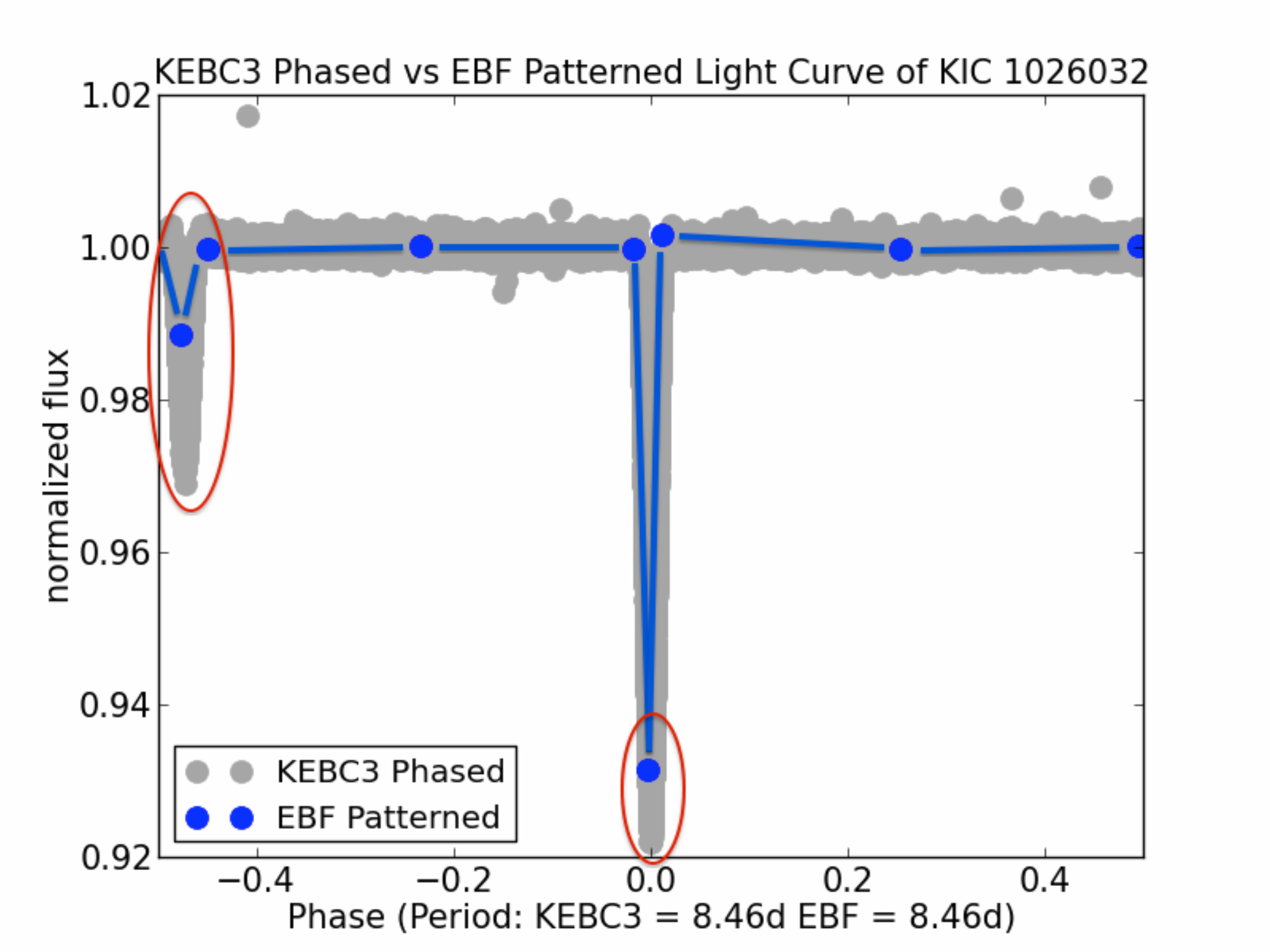}
\caption{\footnotesize{\textbf{Left:} KEBC3 light curve vs. EBF phased, binned light curve of KIC 1026023 detrended by sigma-clipping over $(-1\sigma, 3\sigma)$ interval of the \textit{Legendre} function $P_{10}(x)$ with 200 phase bins. \textbf{Right} KEBC3 light curve vs. EBF patterned curve of KIC 1026032 fit to polynomial chains of no less than three bins. Here the EBF recovers the cataloged EC in spite of failure to capture the full depth of the eclipses; circled in red.}\label{Figure14}}
\end{figure}

The PGM fitted the EBF phased, detrended light curves with either four- or two-polynomial chains and a goodness of fit described by a median $\chi_T^2$ of 5.39 as compared to the unaltered \texttt{polyfit} algorithm used to fit the KEBC3 samples with a median $\chi_T^2$ of 3.74. This goodness of fit includes cases where the PGM patterned fit is in excellent agreement with the manually determined KEBC3 light curves, such as the target shown in Figure \ref{Figure13}, as well as instances where the patterned fit fails to capture to full depth of the eclipses as circled in Figure \ref{Figure14}. 


The subtle deviation between a phased light curve and an EBF generated pattern may present a challenge to an automated solution estimator, yet the PGM's representation of the light curves shape allows the ANN to successfully recover and classify these systems. This and other deviations, such as the small aberrant $3^{rd}$ chain circled in red on Figure \ref{Figure11}, are the result of the \textit{three data points per chain minimum} parameter we adopted in the PGM in order to capture narrow eclipses. Though $\sim$10\% of the non-recovered KEBC3 benchmark sample is comprised of wide EDs with narrow eclipses that contain less than three binned data points, the three data point rule remains a hard minimum to ensure pattern generation integrity. It is important to note that the PGM failed to pattern 3.33\% of the benchmark sample and that these light curves were not considered for follow on recognition by the ANN.

\subsubsection{Classification Metrics from the ANN and CVM \label{resc}}

The performance of the ANN for cataloged binaries in the \textit{Kepler} ``Q3'' data set is benchmarked by KEBC3 EBs with period $p$ where $0.11d\le$ $p$ $\le20.10d$ while ETVs, HBs, and UNCs were excluded. The ANN reports each of its classifications with a posterior probability $P$ interpreted as a level of confidence (LoC) for the morphological class of EC, ESD, ED, or any other periodic variable class contained in the ASAS training set \citep{Paegert2014}. 

Before discussing the performance of the EBF pipeline as a whole, we consider the ANN performance alone. This is evaluated only for those light curves passed forward by the DCM, HRM, and PGM such that the ANN performance metrics are not hindered by the exclusion of light curves by the preceding modules. The ANN's classification accuracy metric is calculated using the KEBC3 targets flagged in the KEBC3 database as InCat:False (ICF), a false positive EB (e.g., blends, planets, pulsators, rotators). Thus we only select those targets designated ICF that lack a cataloged morphology (e.g., pulsators, rotators) as an inaccurate classification. That is, the ANN's accuracy is given as the percentage of ANN EC, ESD, ED classifications that are \textit{Kepler} targets currently cataloged in the KEBC3 as an EB as opposed to those currently cataloged as a pulsating or rotating variable. The ANN's completeness metric is simply the percentage of the adjusted KEBC3 benchmark sample that the ANN successfully recovered as an EB. Using the \textit{Kepler} ``Q3'' data set, the ANN accuracy is found to be 93.68\%, 84.27\%, and 83.03\% with a completeness of 94.17\%, 93.62\%, and 93.63\% for EC, ESD, and ED respectively. 

The CVM's performance was evaluated by a manual inspection of the 91 high-confidence ($P\ge0.90$) EB candidate light curves generated by the EBF. Here we remove from these candidates 21 \textit{Kepler} ``Q3'' targets that suffer from the phasing ambiguity discussed in section \ref{resp}. This manual inspection of the remaining 68 candidates found that the high-confidence CVM validated EBs are likely reliable to $\sim90$\% accuracy, commensurate with the integrated EBF pipeline's accuracy as discussed below.

\subsubsection{Overall Performance}

To evaluate the performance of the integrated EBF pipeline, we describe the EBF generated catalog for candidate ECs, ESDs, and EDs with two distinct queries in order to evaluate EBF accuracy and completeness in the \textit{Kepler} FoV both at high-confidence ($P\ge0.90$) and with no regard to confidence ($P\ge0.01$). Here, \textit{no regard to confidence} means either the primary ANN classification is EC, ESD, ED, or the primary classification is the placeholder MISC with a secondary classification of EC, ESD, or ED with any posterior probability on the interval (0,1). The accuracy and completeness for each condition are listed in Table \ref{Table1}.

\begin{table}[!hb]
\begin{center}
\caption{EBF Pipeline Accuracy and Completeness with the \textit{Kepler} ``Q3'' Data Release.}
\label{Table1}
\begin{tabular}{lcrrrrrr}
\tableline
\tableline
Query & & & EC & ESD & ED &\\
\tableline
\tableline
$P\ge0.90$ & for &$0.11d\le period \le 20.10d$ & & & &\\
 & & Accuracy (\%) & 96.12 & 95.53 & 91.93 &\\
 & & Completeness (\%) & 63.93 & 45.80 & 32.40 &\\
 \tableline
$P\ge0.01$\tablenotemark{*} & for & $0.11d\le period \le 20.10d$ & & & &\\   
 & & Accuracy (\%) & 93.68 & 84.27 & 83.03 &\\
 & & Completeness (\%) & 86.07 & 73.95 & 61.59 &\\
\tableline
\tableline
\end{tabular}
\end{center}
\tablenotetext{*}{Classifications with no regard to confidence. Either the primary ANN classification is EC, ESD, ED, or the placeholder MISC with a secondary classification of EC, ESD, or ED with any posterior probability on the interval (0,1).}
\end{table}

These two queries serve to exemplify the range of utility in the EBF's automated output of candidate \textit{Kepler} FoV EBs. For example, a high-confidence ($P\ge0.90$) query for ECs in the interest of analyzing the characteristics of a homogeneous sample of \textit{Kepler} ECs, with a high certainty of morphological classification, recovers $\sim64\%$ of those available in the ``Q3'' data set with only $\sim4\%$ false positives. Conversely, a query with no regard to confidence ($P\ge0.01$) in the interest of a statistical analysis using a large number of \textit{Kepler} ECs, without regard to the certainty of their morphology ($P\ge0.01$), recovers $\sim86\%$ of those in the ``Q3'' data with $\sim6\%$ false positives.

\subsection{New High Confidence Candidate EBs Identified with the EBF}

In addition, for the high-confidence condition we find 68 CVM validated EB candidates that do not appear in the KEBC3 catalog. The set of these newly identified candidates are provided in Appendix \ref{APP}. In general, these newly identified candidates are in many cases visually more complex and less obviously identified as EBs via manual visual inspection of the light curves. This is not surprising, as the EBs previously cataloged in the KEBC3 were by definition those EBs that could be readily identified by visual human analysis, and therefore these new candidates are only those that would have been most likely to be missed by human visual inspection. Nonetheless, visual inspection of these new candidates identified in an automated manner by the EBF indicates that most if not all of these are bona fide EBs. In any case, these new candidates will be examined further in a follow-on paper.

\section{Discussion \label{DIS}}

Previous works have developed aspects of an automated EB classifier, although to date a fully automated end-to-end pipeline has remained an open challenge. For example, the EB identification algorithm of \citet{Wyrzykowski2003} automatically identified candidates in the Optical Gravitational Lensing Experiment (OGLE), but then relied on visual inspection of all candidate light curves for manual classification. Similarly, the semi-automated method of \citet{Graczyk2011} relied on visual inspection to manually remove artifacts from the OGLE-III candidate light curves. Though the pipeline of \citet{Devor2008} automatically filtered $97\%$ of the Trans-Atlantic Exoplanet Survey (TrES) data as non-periodic, a manual removal of $86\%$ of the remaining light curves was necessary to finalize the catalog. More advanced algorithms described by \citet{Devinney2012}, such as the Eclipsing Binaries via Artificial Intelligence (EBAI) method \citep{Prsa2008}, used sophisticated ANNs to estimate the characteristic parameters of detached EBs in the \textit{Kepler} Field; however, the classification of EB morphological sub-classes remained largely a manual task.

In contrast to the algorithm of \citet{Wyrzykowski2003}, where the authors are only able to estimate that their EB catalog is complete to $\sim85\%$ with an efficiency of $\sim95\%$, and where the classification of EB sub-classes is achieved only through visual inspection, the EBF's completeness and accuracy rates are precisely benchmarked by the KEBC3 and based upon the ANN's fully-automated classification of sub-classes. Additionally, the EBF's single-layer ANN represents a much simpler representation of the classifier algorithm with a more robust training set (i.e., $19$ input nodes and trained on $\sim32,000$ distinct ASAS exemplars). The \citet{Wyrzykowski2003} method requires the conversion of OGLE photometric light curves into representational pixel images that are then analyzed by a $1,050$ input node three-layer ANN trained on variations of only $10$ OGLE EB exemplars.

The OGLE-III analysis of \citet{Graczyk2011} is based on a $13$ step algorithm that, when complete, reports a false positive rate of $\sim20\%$ and thus requires a manual inspection of each and every EB candidate light curve. In contrast, the EBF's automated filter reduces false positives to less than $\sim10\%$. The semi-automatic method of \citet{Devor2008} also requires manual classification of the EB sub-classes and is further constrained to the analysis of ED and ESD TrES light curves while the EBF extends the result to include the EC class. The EBF pipeline again represents an advance in full automation and adds the ability to automatically classify the EC sub-class when compared to more advanced solution estimators, such as the EBAI method \citep{Prsa2008}.

The purpose of the EBF pipeline is to alleviate the human bottleneck in discovering astrophysically interesting eclipsing binary systems by automatically identifying and classifying these systems from large volumes of survey data as a foundation for an IDP designed for the future of astroinformatics. 
The performance results using the \textit{Kepler} ``Q3'' Data Release demonstrate the EBF's accuracy for those EBs flagged at high confidence, however the completeness metric leaves room for improvement. An examination of the pipeline's unclassified KEBC3 EBs reflects the 5.59\% exclusion rate of the AoV-SNR output filter, as well as the 3.33\% failure rate of the PGM. Here we find that 8.92\% of the benchmark sample never reaches the ANN. In addition, we discover that a simple evaluation of the flux ratio in relation to an arbitrary threshold of $\sim0.5$ is insufficient to discriminate between ECs and other variable types. Thus, the necessity for the pipeline to first reliably identify harmonics as well as correctly phase and fit the input data without manual adjustment is obviously an area in need of refinement. 

These period and subsequent phasing errors can potentially be mitigated in the future by layering the HRM with multiple harmonic search methods for varied period length. For example, the computationally inexpensive Fast Chi Squared (F$\chi^2$) method of \citet{Palmer2009} and the Boxed Least Squares (BLS) transit search algorithm \citep{Kovacs2002} may supplement the AoV method. This multi-layered harmonics search could be combined with a revised PGM where a post-fitting $\chi^2$ analysis may be used to detect errors and validate assigned periods. Phasing may then be subjected to other constraints (e.g., the symmetry vs asymmetry of the primary eclipse ingress and egress for those targets with FR $\sim0.5$). Training deficits such as the lack of a sufficient number of wide EB light curves with narrow transits in the ASAS exemplars as well as an inherent morphological ambiguity in binary systems can potentially be rectified with further training, testing, and validation using expanded survey data sets. 

The EBF's adaptability, facilitated by the modular parameters described in Section \ref{MET}, and the automated catalog's query options, demonstrates the pipeline's utility for a variety of applications. Applying the EBF to other current time-domain photometric surveys, such as K2 and the upcoming TESS mission should provide similar results. In addition, the EBF's ANN is currently tooled and trained to recognize and classify the other types of periodic variables discussed in \citet{Paegert2014} (i.e., the ASAS periodic variable classifications DCEP-FU, DCEP-FO, DSCT, RRAB, RRC, and MIRA). Only a benchmark of variables in the \textit{Kepler} FoV, parallel to the KEBC3, is needed to tool and test the EBF's expansion into automatically cataloging these and other types of periodic variables.

\section{Conclusion \label{CON}}

The Eclipsing Binary Factory (EBF) generated a fully automated catalog of eclipsing binary systems in the \textit{Kepler} field of view from the \textit{Kepler} ``Q3'' Data Release. This catalog, benchmarked by the manually generated \textit{Kepler} Eclipsing Binary Catalog v3, identified and classified to 90\% certainty eclipsing contact (EC), eclipsing semi-detached (ESD), and eclipsing detached (ED) systems ($0.11d\le period \le 20.10d$) with an accuracy of 96.12\%, 95.53\%, and 91.93\%, while complete to 63.93\%, 45.80\%, and 32.40\% respectively. When classification certainty is not considered, the EBF catalog identified and classified ECs, ESDs, and EDs ($0.11d\le period \le 20.10d$) with an accuracy of 93.68\%, 84.27\%, and 91.93\%, while complete to 86.07\%, 73.95\%, and 61.59\% respectively. In addition, the EBF identified 68 new candidate EBs that may have been missed by the human generated KEBC3 (see Appendix \ref{APP}).

This demonstrates the EBF pipeline as an effective alternative to the constraints of human analysis when probing large time-domain, photometric data sets. As such, the EBF computational pipeline is a viable framework for the development of a more complex intelligent data pipeline containing a follow on automated solution estimator that would enable the automatic determination of physical EB parameters. Future work on the EBF will include enhancing the harmonics identification capabilities with multiple search methods and phasing constraints beyond the flux ratio, a revised pattern generator to constrain polynomial chain fits to smaller errors, and more robust training for the ANN on wide binary systems. In addition, an expansion of EBF searches in the \textit{Kepler} data, and in the upcoming K2 data, to include the periodic variables of type $\delta-$cepheid (fundamental and first overtones), $\delta-$scuti, RR-Lyrae ($ab$ and $c$), and Mira will be published in follow-up papers. Over the long term, the EBF general classification approach can be adopted by future surveys such as TESS and LSST.

\acknowledgments
We are grateful to A. Pr\v{s}a, K. Conroy, and all those at Villanova University who contribute to
the \textit{Kepler} Eclipsing Binary Catalog, to J. Pepper for his insights and experience with large surveys, and to R. Siverd for his expert assistance with harmonics.
Kepler was selected as the $10^{th}$ mission of the Discovery Program. Funding for this mission is provided
by NASA, Science Mission Directorate.

\pagebreak

\appendix

\section{EBF High-Confidence New EB Candidates\label{APP}}

\begin{figure}
\begin{center}
\includegraphics[scale=0.5]{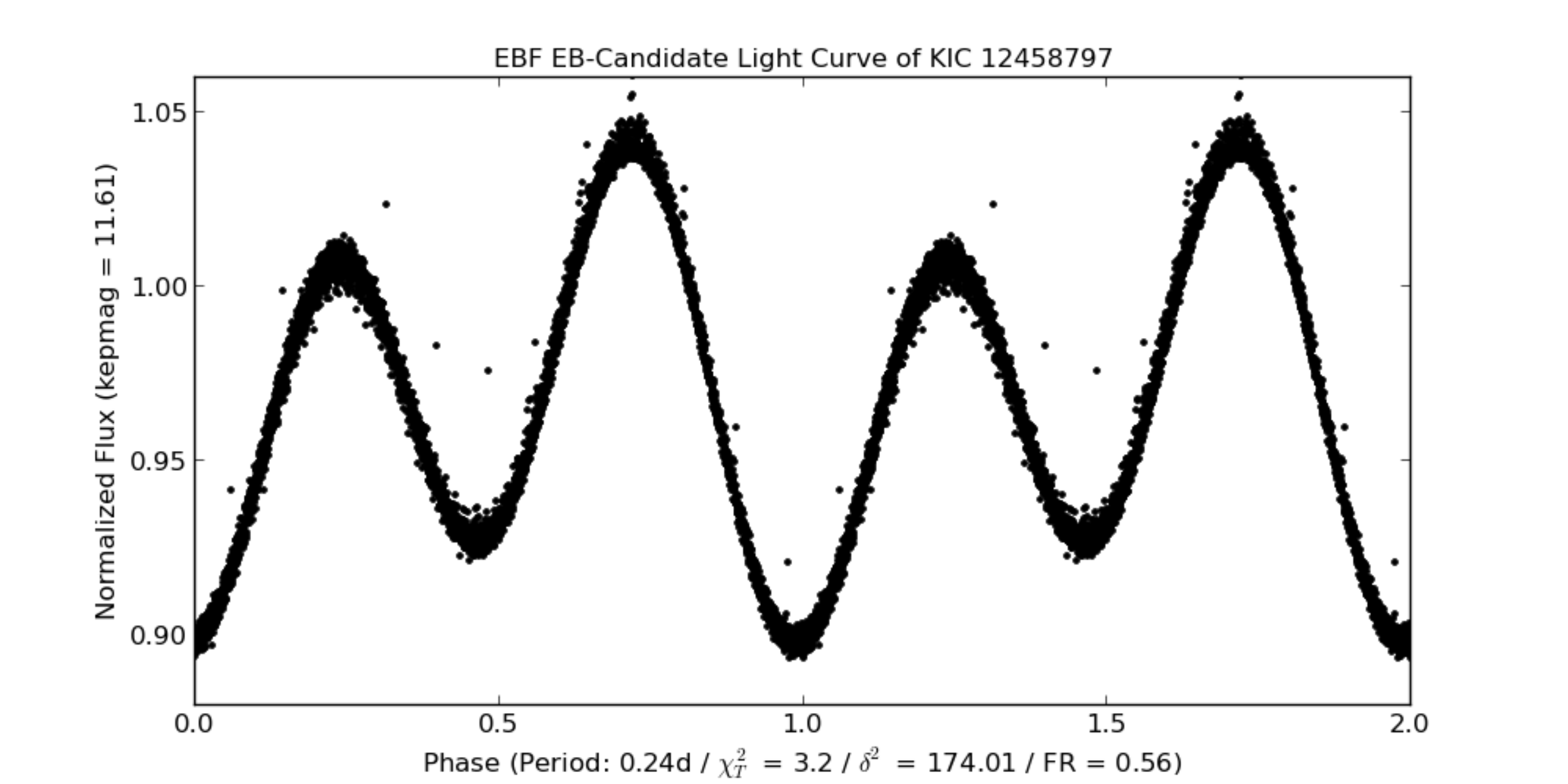}
\includegraphics[scale=0.5]{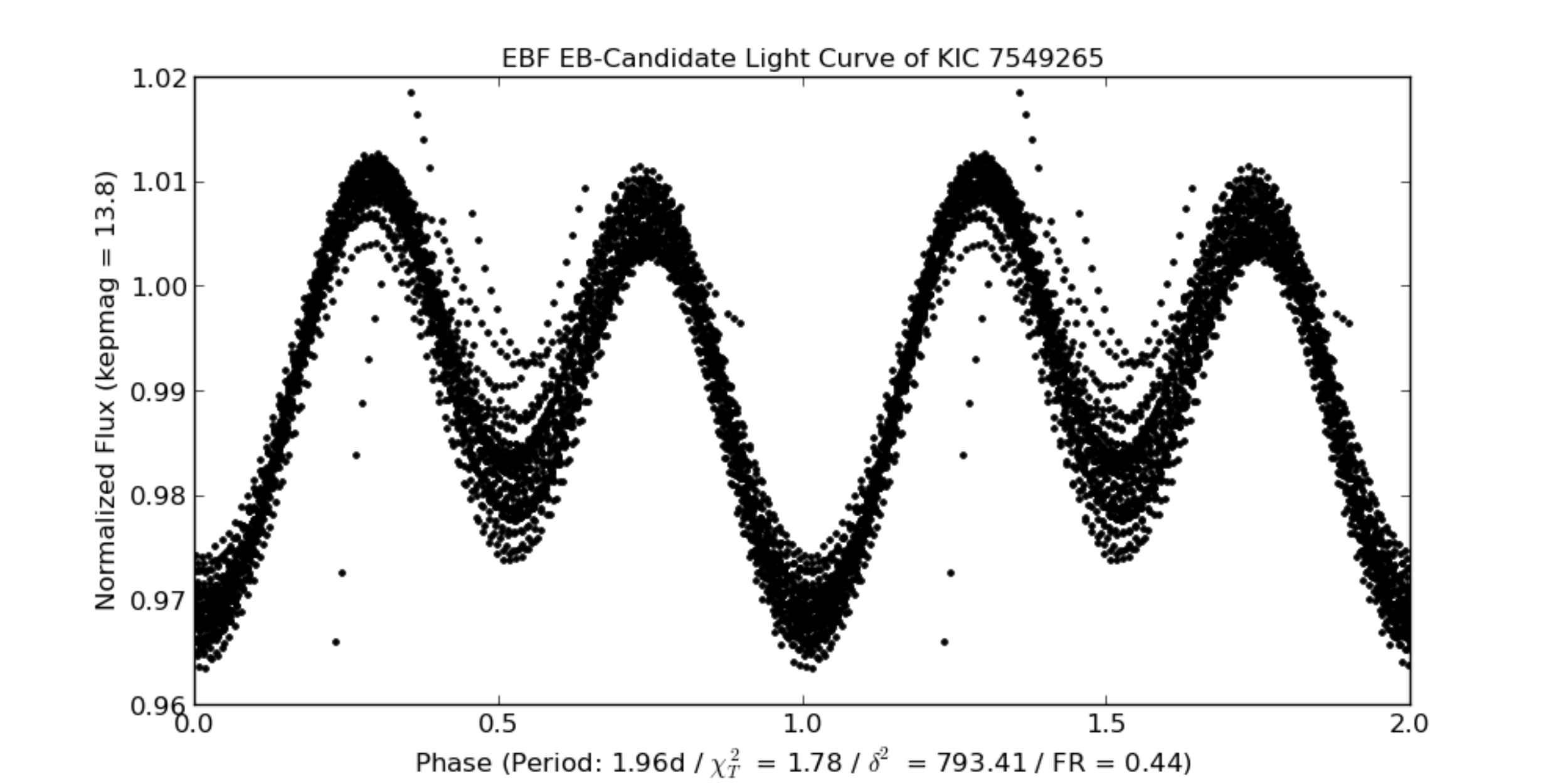}
\includegraphics[scale=0.5]{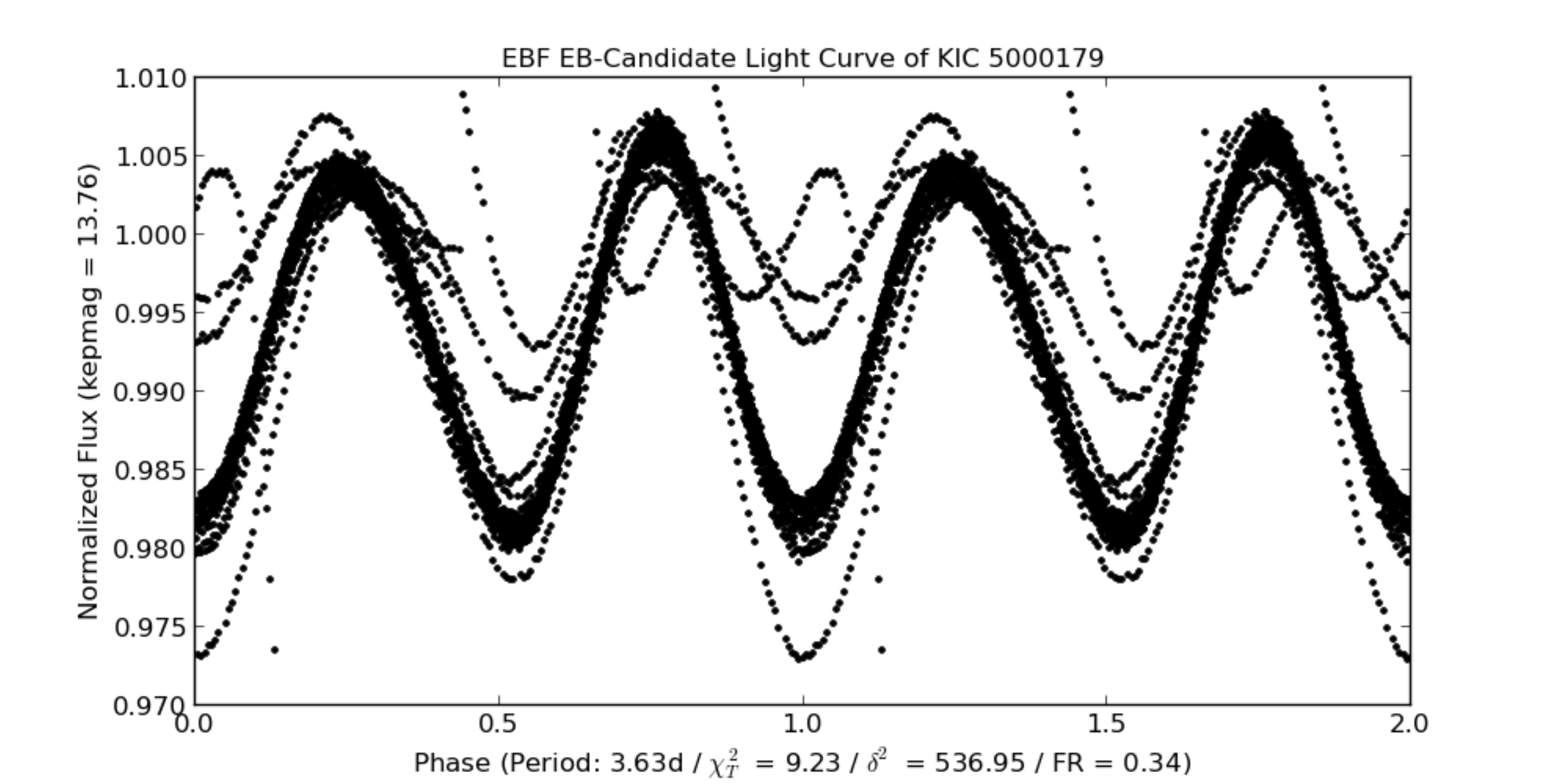}
\caption{\footnotesize{EBF generated \textit{Kepler} ``Q3'' EB candidate light curves for KIC 12458797, KIC 7549265, and KIC 5000179}\label{Figure15}}
\end{center}
\end{figure}

\begin{figure}
\begin{center}
\includegraphics[scale=0.5]{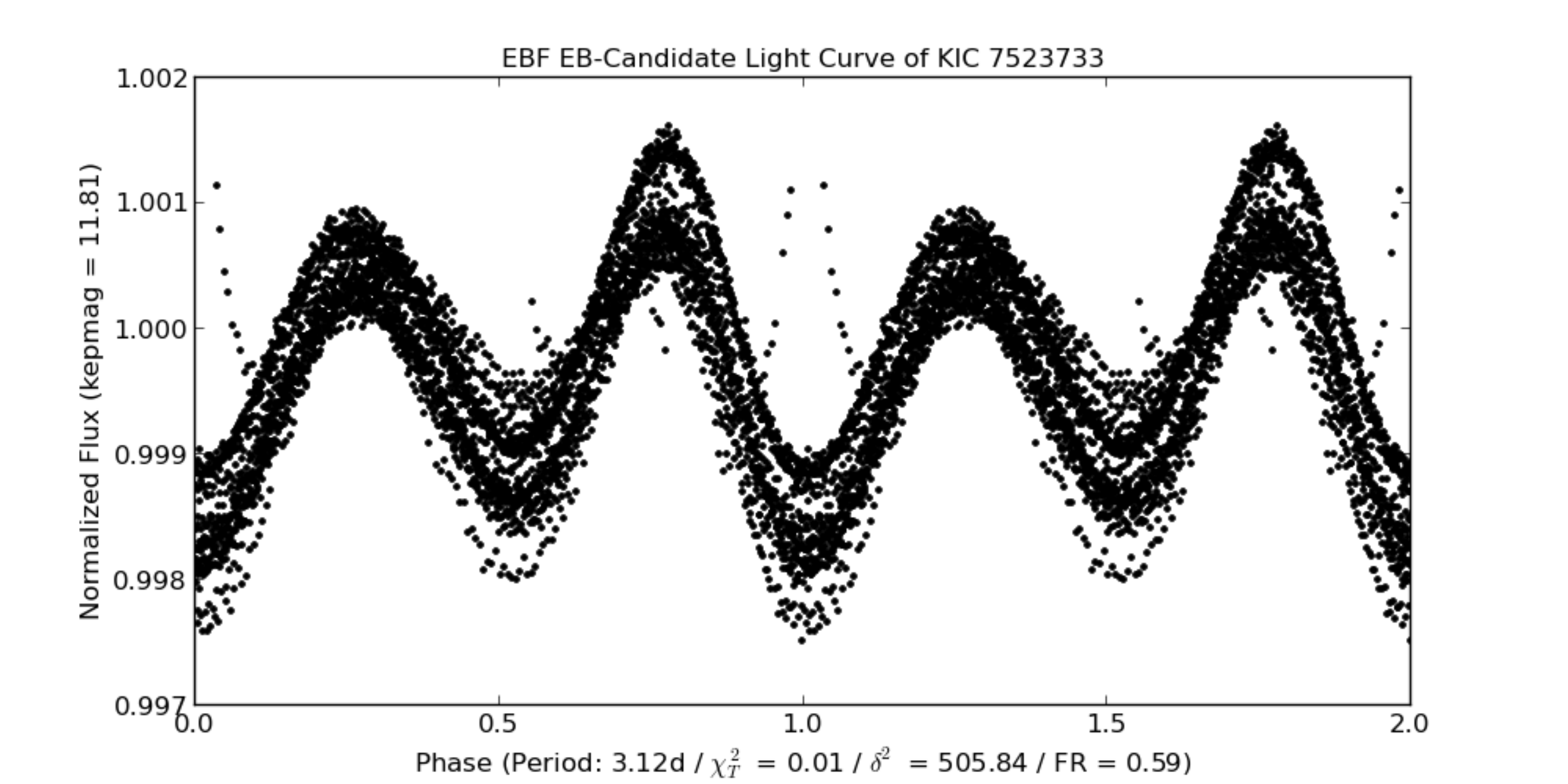}
\includegraphics[scale=0.5]{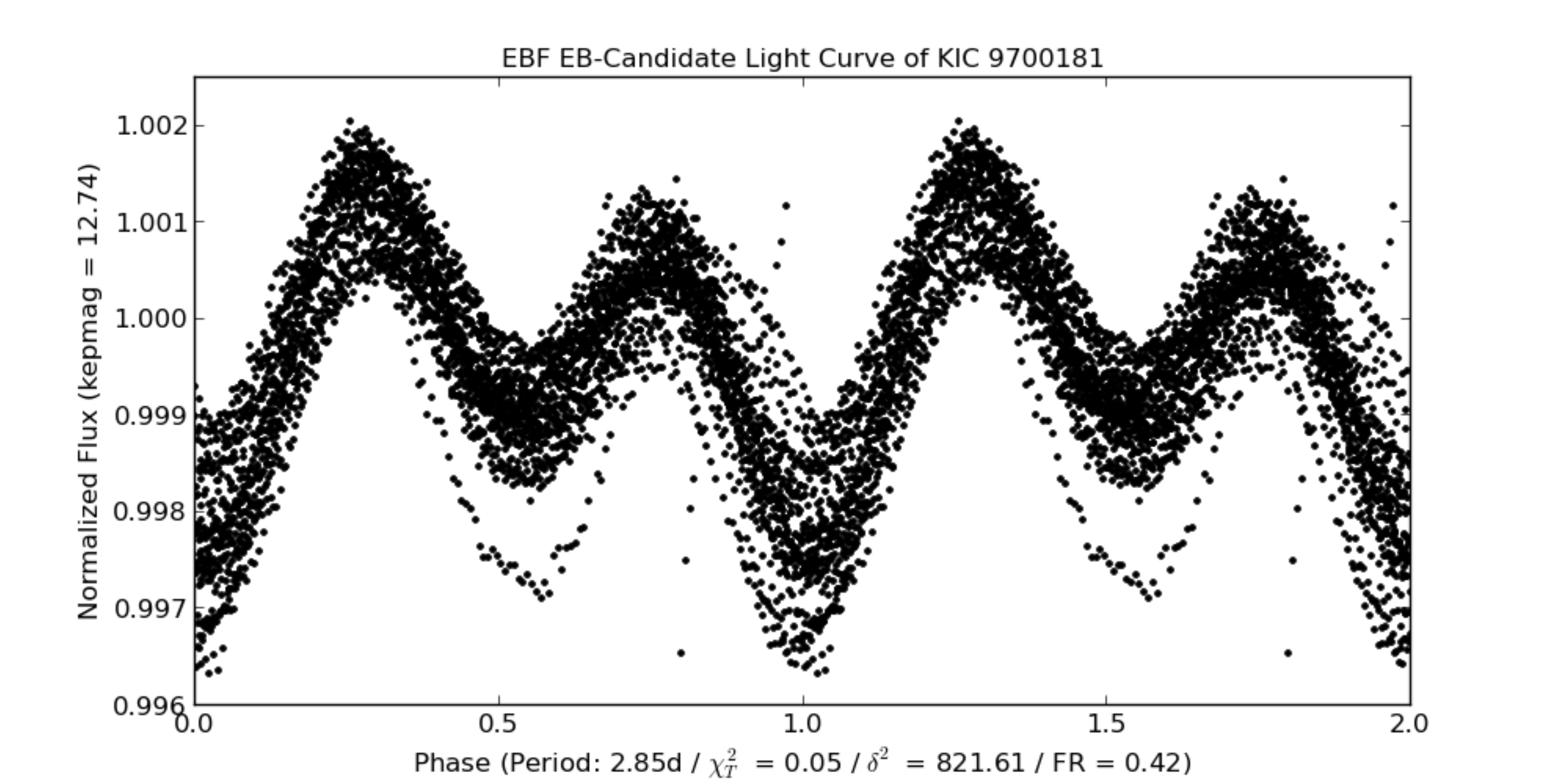}
\includegraphics[scale=0.5]{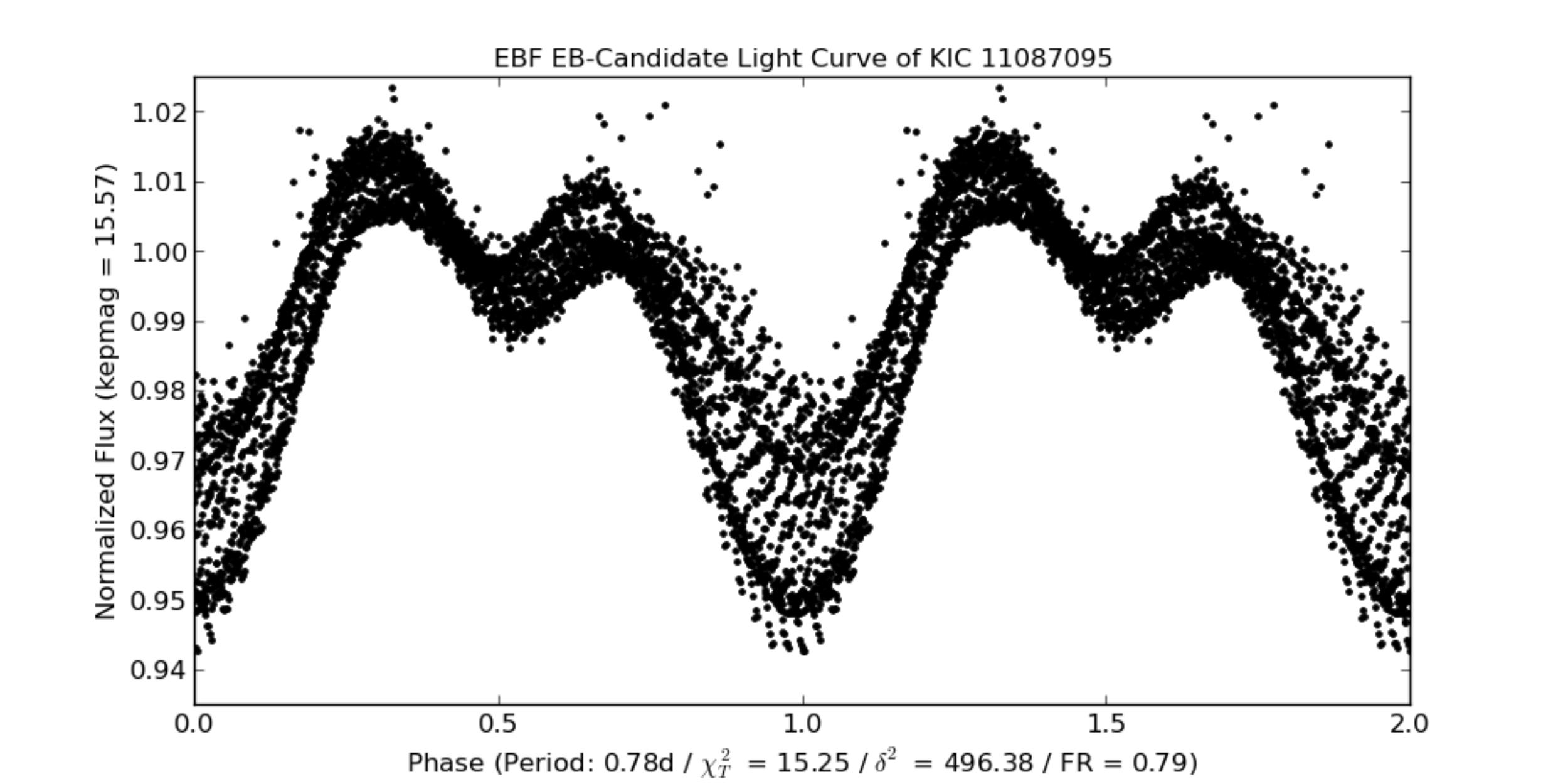}
\caption{\footnotesize{EBF generated \textit{Kepler} ``Q3'' EB candidate light curves for KIC 7523733, KIC 9700181, and KIC 11087095}\label{Figure16}}
\end{center}
\end{figure}

\begin{figure}
\begin{center}
\includegraphics[scale=0.5]{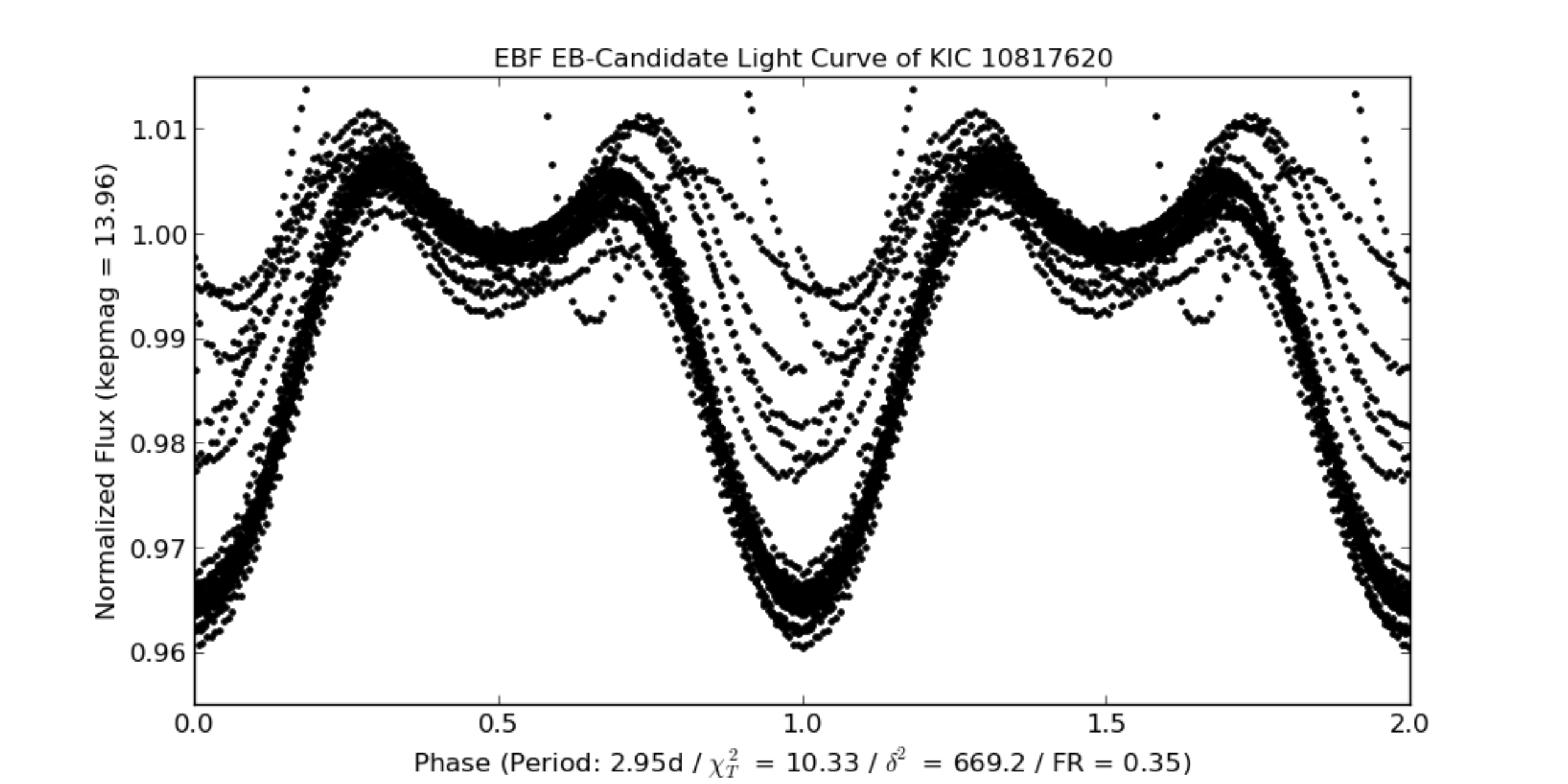}
\includegraphics[scale=0.5]{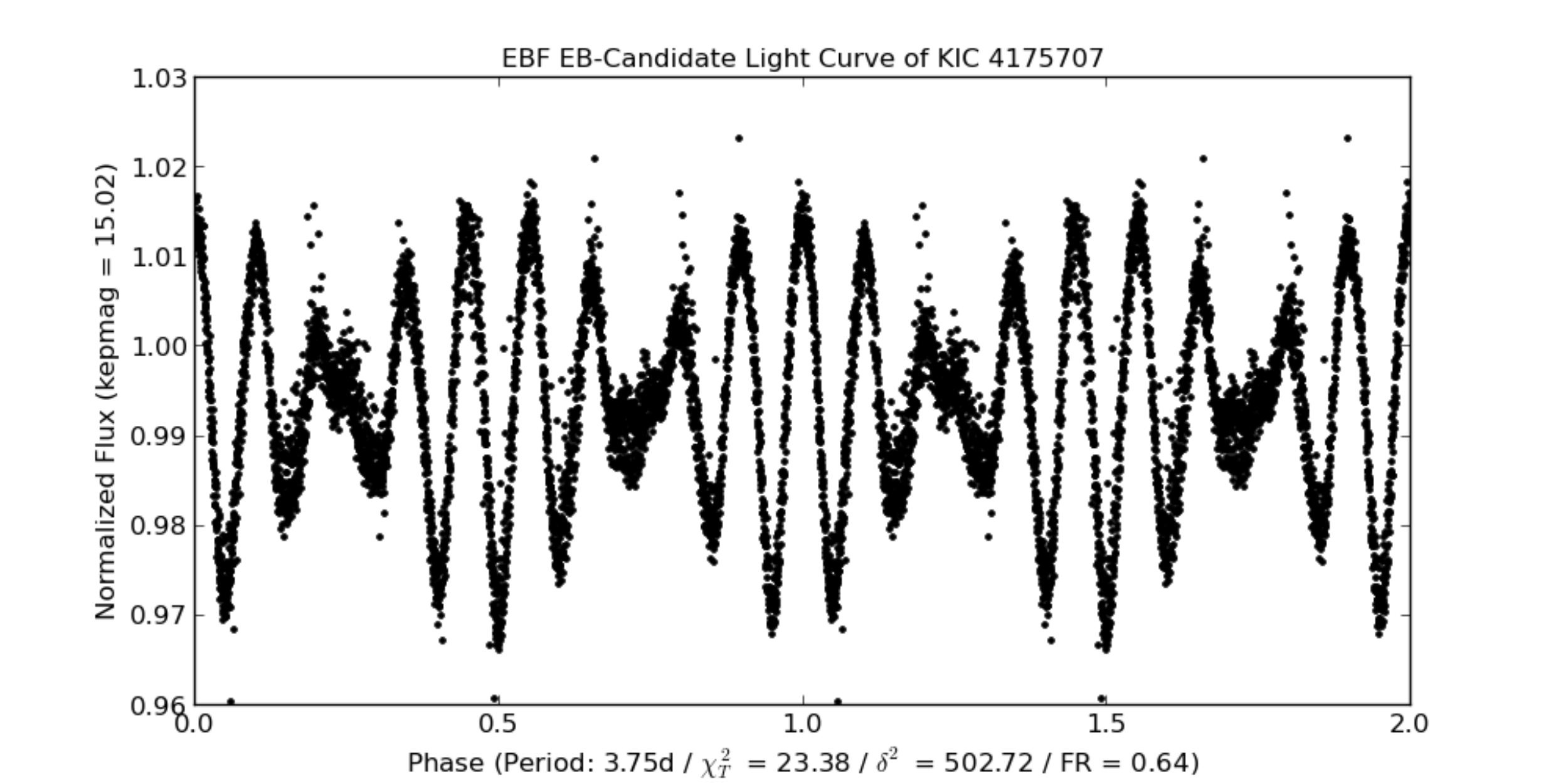}
\includegraphics[scale=0.5]{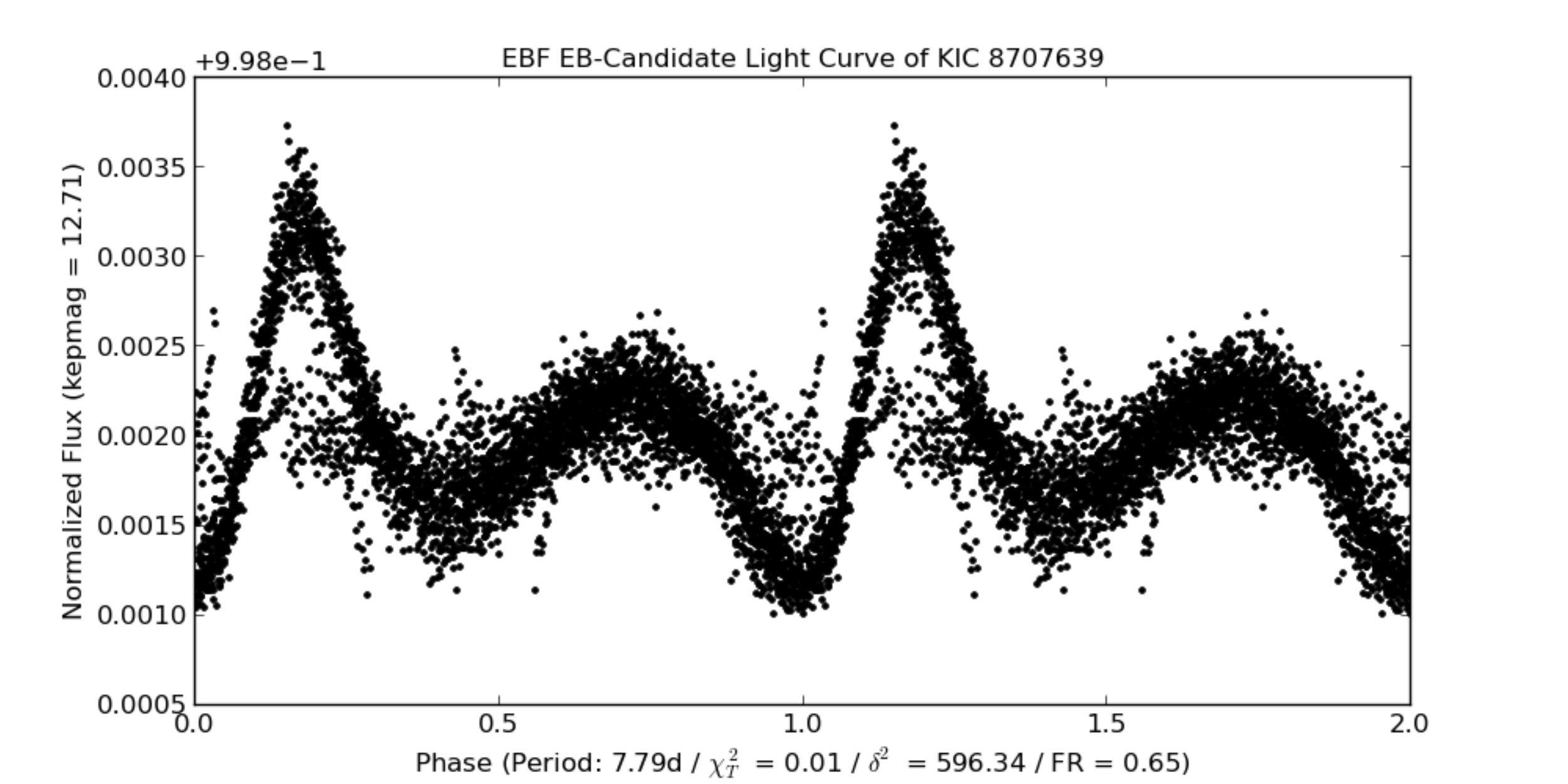}
\caption{\footnotesize{EBF generated \textit{Kepler} ``Q3'' EB candidate light curves for KIC 10817620, KIC 4175707, and KIC 8707639}\label{Figure17}}
\end{center}
\end{figure}

\begin{figure}
\begin{center}
\includegraphics[scale=0.5]{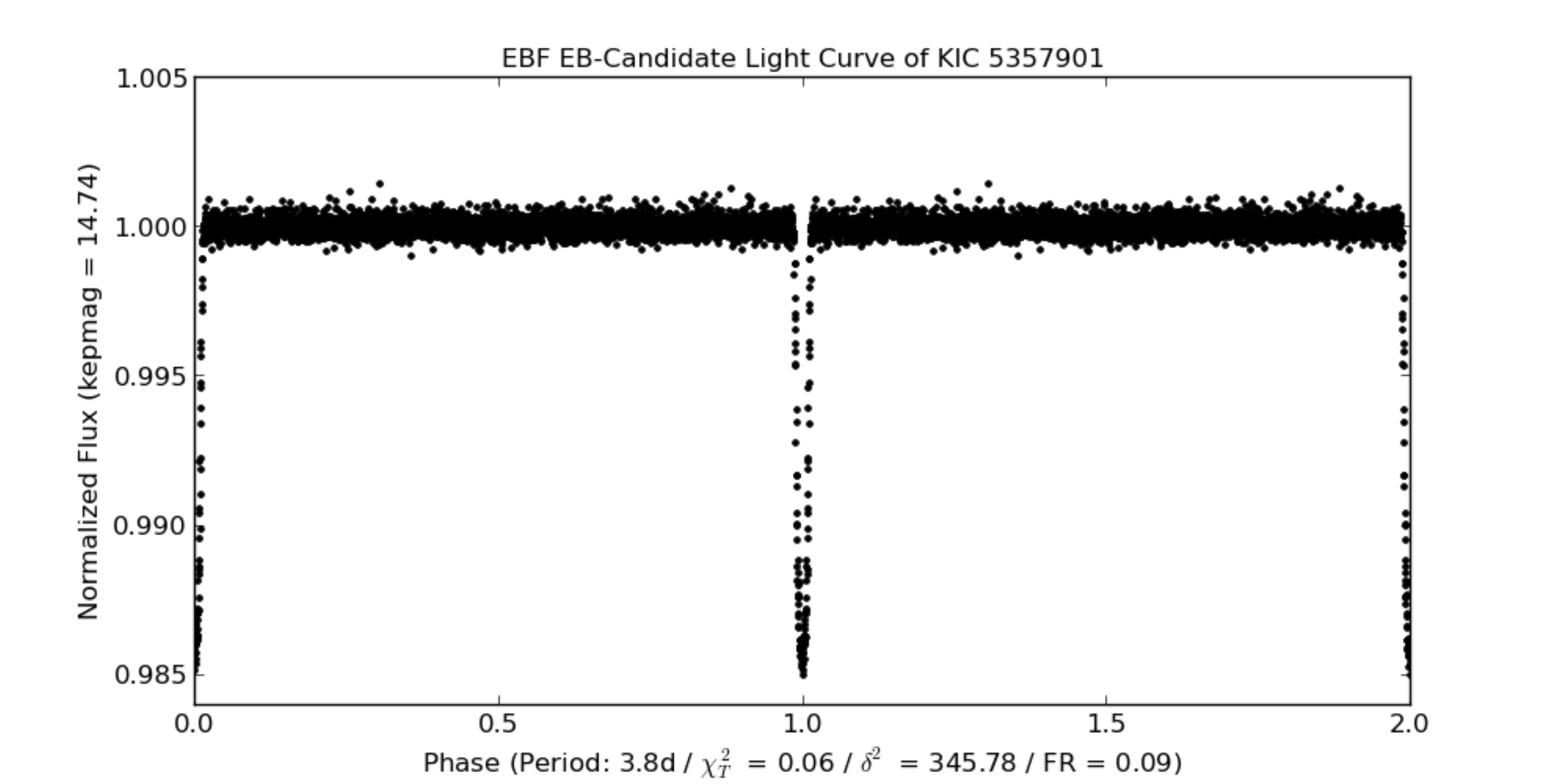}
\includegraphics[scale=0.5]{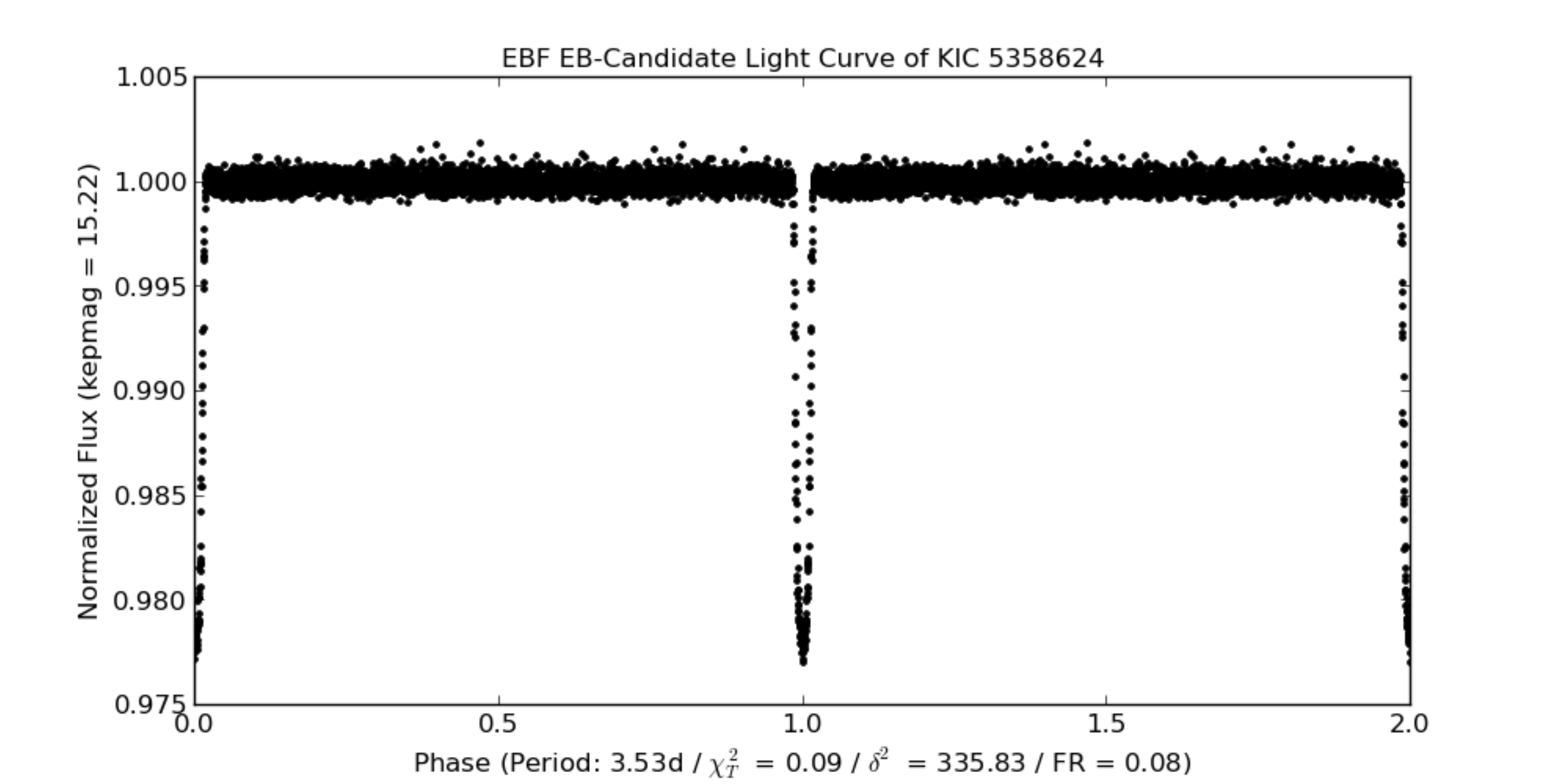}
\includegraphics[scale=0.5]{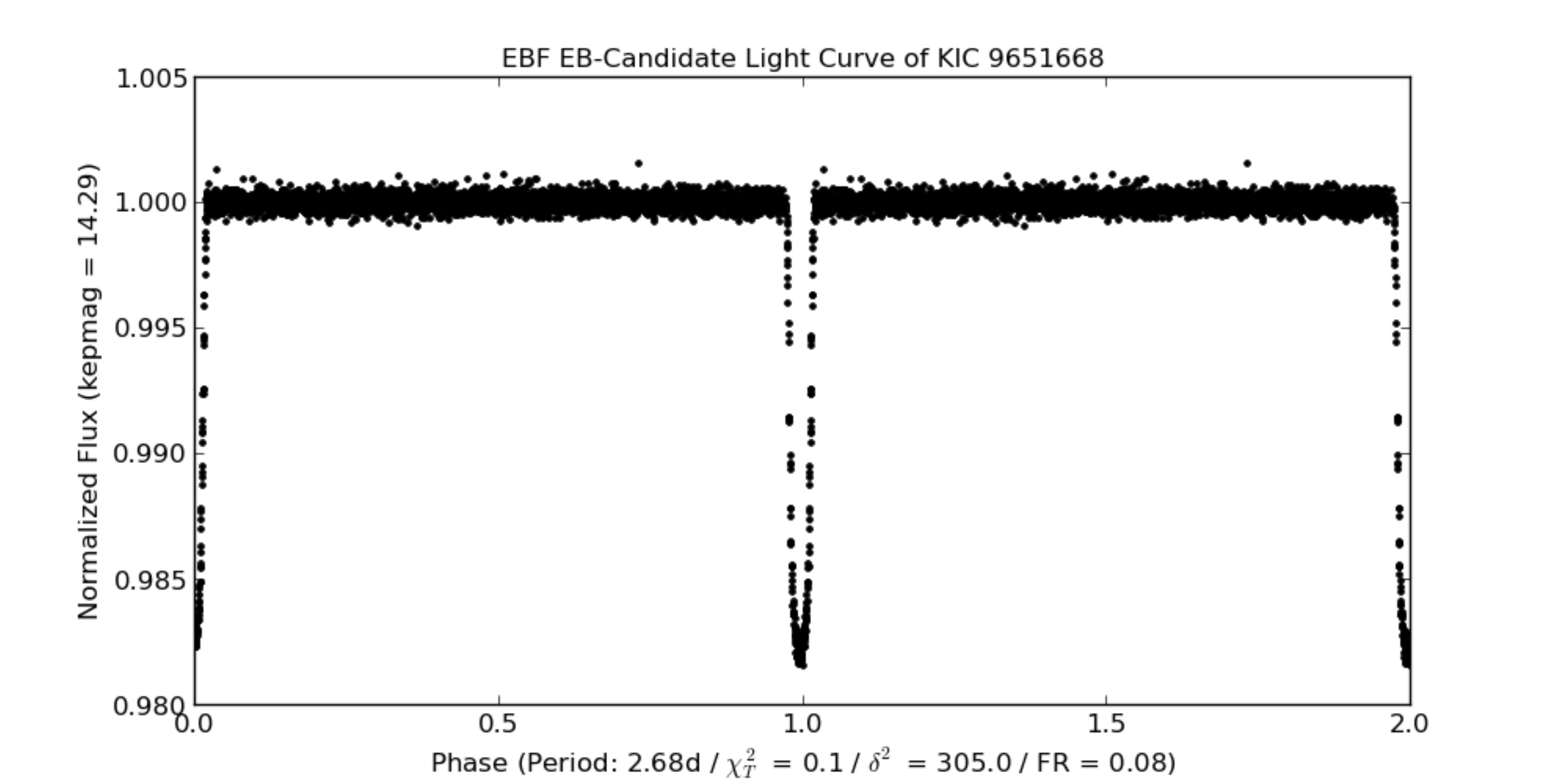}
\caption{\footnotesize{EBF generated \textit{Kepler} ``Q3'' EB candidate light curves for KIC 5357901 (KOI K00188.01), KIC 5358624 (KOI K00830.01), and KIC 9651668 (KOI K00183.01) each with the current disposition of ``CANDIDATE'' in the \textit{Kepler} Objects of Interest (KOI).}\label{Figure18}}
\end{center}
\end{figure}

In Table \ref{Table2}, we provide the KIC, RA, DEC, period, and \textit{Kepler} magnitude for 68 high-confidence ($P\ge0.9$), astrophysically interesting EB candidates newly identified by the EBF pipeline from the Kepler ``Q3'' light curves that are not currently included in the human generated \textit{Kepler} Eclipsing Binary Catalog v3.
Figures \ref{Figure15}--\ref{Figure17} represent a sample of ``Q3'' candidate light curves. The list of candidates also includes three targets with a current disposition of ``CANDIDATE'' in the \textit{Kepler} Objects of Interest (KOI). Figure \ref{Figure18} shows the light curves for KOI K00188.01, K00830.01, and K00183.01.

\begin{table}[!hb]
\begin{center}
\caption{High-confidence ($P\ge0.9$) candidate EBs newly identified by the EBF pipeline from the Kepler ``Q3'' light curves that are not currently included in the human generated \textit{Kepler} Eclipsing Binary Catalog v3.}
\label{Table2}
\begin{tabular}{|lllll|}
\tableline
KIC & RA & DEC & Period (d) & KEPMAG\\
\tableline
1872192 & 292.29273 & 37.32021 & 0.669979 & 13.74\\
2164791 & 292.427369 & 37.556019 & 6.703367 & 15.22\\
2285420 & 286.776414 & 37.63955 & 5.347988 & 13.5\\
2696217 & 286.90421 & 37.90544 & 12.807157 & 13.45\\
3441414 & 290.84053 & 38.54963 & 8.114992 & 11.51\\
3761175 & 295.0293 & 38.89788 & 5.162696 & 12.83\\
3831911 & 285.05514 & 38.98501 & 7.339436 & 15.02\\
3836276 & 286.88712 & 38.90174 & 8.176842 & 15.32\\
3847822 & 290.738025 & 38.971723 & 7.256877 & 11.85\\
4164363 & 293.210621 & 39.2407 & 5.791975 & 11.37\\
4175707 & 295.698821 & 39.26685 & 3.746223 & 15.02\\
4271063 & 293.72196 & 39.38955 & 0.989814 & 14.11\\
4451525 & 287.60961 & 39.59948 & 16.264671 & 13.88\\
4926962 & 292.68917 & 40.04848 & 4.463288 & 13.13\\
5000179 & 288.35829 & 40.17466 & 3.632168 & 13.76\\
5083330 & 286.319321 & 40.23009 & 7.814972 & 15.73\\
5083911 & 286.585518 & 40.228043 & 12.133361 & 12.96\\
5357901 & 290.35802 & 40.56774 & 3.797303 & 14.74\\
5358624 & 290.58173 & 40.5774 & 3.525784 & 15.22\\
5391520 & 298.830296 & 40.507433 & 3.984854 & 11.93\\
5857714 & 284.69325 & 41.16556 & 0.163569 & 11.21\\
6207355 & 292.59219 & 41.53943 & 5.210091 & 13.46\\
6425891 & 285.25158 & 41.8777 & 0.951153 & 14.77\\
6507888 & 286.7031 & 41.99742 & 7.453959 & 15.27\\
6613627 & 293.97651 & 42.04829 & 0.226224 & 12.55\\
6616211 & 294.67262 & 42.01478 & 3.85509 & 11.79\\
6924320 & 282.394541 & 42.44624 & 16.505877 & 15.49\\
6953069 & 293.20455 & 42.48449 & 0.294107 & 15.74\\
7053456 & 297.03561 & 42.53181 & 11.211459 & 12.0\\
7219251 & 296.87916 & 42.70233 & 3.190739 & 13.84\\
7363684 & 292.06889 & 42.93308 & 0.729569 & 13.72\\
7431838 & 287.45459 & 43.04541 & 3.036295 & 12.09\\
\tableline
\end{tabular}
\end{center}
\end{table}

\begin{table}[!hb]
\begin{center}
\begin{tabular}{|lllll|}
\tableline
KIC & RA & DEC & Period (d) & KEPMAG\\
\tableline
7434110 & 288.35027 & 43.04639 & 7.223437 & 15.99\\
7523733 & 290.58686 & 43.11656 & 3.121057 & 11.81\\
7549265 & 297.31412 & 43.14078 & 1.962315 & 13.8\\
7748449 & 290.22456 & 43.49293 & 15.705378 & 13.43\\
7905209 & 296.36599 & 43.62922 & 0.209777 & 13.27\\
8029556 & 292.09914 & 43.88178 & 18.040183 & 15.35\\
8098181 & 292.01466 & 43.97829 & 0.986971 & 12.86\\
8432040 & 292.30794 & 44.48024 & 7.991428 & 11.95\\
8520065 & 299.48401 & 44.50633 & 9.365808 & 13.99\\
8694772 & 293.97896 & 44.85794 & 15.187454 & 14.83\\
8707639 & 297.887694 & 44.868264 & 7.785182 & 12.71\\
8774912 & 298.697591 & 44.90004 & 15.137412 & 12.99\\
8881943 & 290.48391 & 45.18651 & 7.777886 & 14.79\\
8972908 & 297.87374 & 45.24537 & 4.195616 & 12.58\\
9011963 & 287.87358 & 45.31096 & 8.400634 & 15.5\\
9283572 & 292.93673 & 45.73978 & 1.217898 & 14.36\\
9511937 & 284.28423 & 46.11628 & 0.277782 & 15.54\\
9651668 & 292.85574 & 46.39118 & 2.68435 & 14.29\\
9700181 & 286.87755 & 46.43377 & 2.846435 & 12.74\\
9726020 & 297.60386 & 46.48611 & 10.642508 & 12.94\\
10389596 & 284.504783 & 47.550999 & 2.608079 & 13.36\\
10669516 & 293.44307 & 47.98749 & 10.485432 & 13.03\\
10817620 & 298.71137 & 48.13449 & 2.946091 & 13.96\\
10982373 & 294.90293 & 48.47721 & 10.01776 & 11.7\\
10990083 & 297.79189 & 48.49364 & 10.715948 & 13.24\\
11087095 & 293.33433 & 48.69283 & 0.77833 & 15.57\\
11152786 & 298.17105 & 48.7121 & 8.229868 & 15.41\\
11236035 & 287.51454 & 48.93461 & 2.318711 & 14.65\\
11284185 & 284.144111 & 49.036 & 15.705099 & 13.66\\
11653122 & 286.610501 & 49.79567 & 2.650525 & 14.76\\
11922402 & 295.85559 & 50.28799 & 5.917946 & 15.93\\
12102573 & 286.24793 & 50.61843 & 10.942348 & 11.81\\
12105278 & 288.17649 & 50.67654 & 7.110381 & 12.32\\
12314646 & 295.343081 & 51.07426 & 13.579375 & 14.14\\
12458797 & 290.53101 & 51.30857 & 0.241319 & 13.89\\
12602567 & 290.586697 & 51.602235 & 10.241997 & 13.02\\
\tableline
\end{tabular}
\end{center}
\end{table}
\pagebreak

\end{document}